\documentclass[twocolumn,showpacs,preprintnumbers,amsmath,amssymb,superscriptaddress,nofootinbib,english]{revtex4}
\usepackage{times,amsmath,amsfonts,amssymb,epstopdf}
\usepackage{graphicx}
\usepackage{dcolumn}
\usepackage{bm}
\usepackage{epsfig}
\usepackage{graphicx}
\usepackage{hyperref}
\usepackage[usenames]{color}
\usepackage{url}
\usepackage[normalem]{ulem}
\usepackage[T1]{fontenc}
\usepackage[dvipsnames]{xcolor}

\newcommand{\remove}[1]{}
\newcommand{\dd}{\mathrm{d}}

\def\be{\begin{equation}}
\def\ee{\end{equation}}

\newcommand{\beq}{\begin{equation}}
\newcommand{\eeq}{\end{equation}}
\newcommand{\beqa}{\begin{eqnarray}}
\newcommand{\eeqa}{\end{eqnarray}}

\renewcommand{\pl}{\partial}

\newcommand{\vx}{{\bf x}}

\renewcommand{\vr}{{\bf r}}

\newcommand{\vF}{{\bf F}}

\newcommand{\tg}{{\tilde{g}}}

\newcommand{\cG}{{\cal G}}

\newcommand{\cM}{{\cal M}}

\newcommand{\Om}{\Omega_{\rm m}}

\newcommand{\bea}{\begin{array}}
\newcommand{\ea}{\end{array}}

\newcommand{\MPl}{M_{\rm Pl}}

\begin{document}
\title{K-mouflage gravity models that pass Solar System and cosmological constraints}

\author{Alexandre Barreira}\email{a.m.r.barreira@durham.ac.uk}
\affiliation{Institute for Computational Cosmology, Department of Physics, Durham University, Durham DH1 3LE, UK}
\affiliation{Institute for Particle Physics Phenomenology, Department of Physics, Durham University, Durham DH1 3LE, UK}
\author{Philippe Brax}\email{philippe.brax@cea.fr}
\affiliation{Institut de Physique Th\'eorique, Universit\'e Paris Saclay, CEA, CNRS, F-91191 Gif-sur-Yvette, France}
\author{Sebastien Clesse}\email{sebastien.clesse@unamur.be}
\affiliation{Namur Center of Complex Systems (naXys), Department of Mathematics, University of Namur, Rempart de la Vierge 8, 5000 Namur, Belgium}
\author{Baojiu Li}\email{baojiu.li@durham.ac.uk}
\affiliation{Institute for Computational Cosmology, Department of Physics, Durham University, Durham DH1 3LE, UK}
\author{Patrick Valageas}\email{patrick.valageas@cea.fr ;  }
\affiliation{Institut de Physique Th\'eorique, Universit\'e Paris Saclay, CEA, CNRS, F-91191 Gif-sur-Yvette, France}


\begin{abstract}

We show that Solar System tests can place very strong constraints on K-mouflage models of gravity, which are coupled scalar field models with nontrivial kinetic terms that screen the fifth force in regions of large gravitational acceleration. In particular, the bounds on the anomalous perihelion of the Moon imposes stringent restrictions on the K-mouflage Lagrangian density, which can be met when the contributions of higher-order operators in the static regime are sufficiently small. The bound on the rate of change of the gravitational strength in the Solar System constrains the coupling strength $\beta$ to be smaller than
$0.1$. These two bounds impose tighter constraints than the results from the Cassini satellite and Big Bang Nucleosynthesis. Despite the Solar System restrictions, we show that it is possible to construct viable models with interesting cosmological predictions. In particular, relative to $\Lambda$-CDM, such models predict percent-level deviations for the clustering of matter and the number density of dark matter haloes. This makes these models predictive and testable by forthcoming observational missions.

\end{abstract}


\maketitle

\maketitle
\section{Introduction}

{\it K-mouflage} is one of the three types of screening mechanisms \cite{Khoury:2013tda}, together with {\it chameleon} and {\it Vainshtein},  for scalar modifications of gravity with a conformal coupling to matter \cite{Clifton:2011jh}. Roughly  speaking, these three mechanisms can be distinguished by what triggers their implementation: K-mouflage operates in regions where the gravitational acceleration is large enough; chameleons \cite{Khoury:2003aq,Khoury:2003rn} are at play when the Newtonian potential is large; and Vainshtein \cite{Vainshtein:1972sx} is active when the spatial curvature becomes significant.  For the cases of K-mouflage and Vainshtein, Newtonian gravity can be preserved deep inside the so-called K-mouflage or Vainshtein radii \cite{Babichev:2009ee,Brax:2012jr}, which are defined as the distances from the center of a spherical source within which the nonlinearities of the scalar field Lagrangian become significant and, therefore, the screening effects come into play. These two mechanisms share in common the fact that the screening efficiency (roughly determined by the size of these radii) depends solely on the properties of the gravitational source, as opposed to chameleons, for which there is also a dependence on the density of the environment where the sources live.

The existence of screening mechanisms, however, gives only the models a chance to pass Solar System tests of gravity \cite{Will:2014xja}. In other words, it is still necessary to investigate further the conditions that make the screening mechanisms efficient enough to cope with the current observational bounds. For instance, even well within the K-mouflage radius, the total force may still exhibit a non-Newtonian component (i.e.~a radial dependence that differs from $1/r^2$), which albeit small, could still be large enough to induce detectable perturbations in the orbits of planets and moons. Currently, Lunar Laser Ranging experiments \cite{2002nmgm.meet.1797W,Gruzinov:2002sa,Dvali:2002vf,Brax:2011sv}, constrain the anomalous perihelion of the Moon at the $10^{-11}$ level, which can then be used to constrain the K-mouflage Lagrangian density term, $K(\chi)$ (cf.~Eq.~(\ref{K-def})). The function $K(\chi)$ is a nonlinear function (to ensure screening) of the kinetic energy term $\chi=-(\partial\varphi)^2/2{\cal M}^4$, where $\varphi$ is the K-mouflage scalar field and ${\cal M}^4$ the dark energy scale. In this paper, we shall use the anomalous perihelion of the Moon to determine the constraints on the shape of $K(\chi)$ for static configurations of $\varphi$. The static regime is that relevant for very small scales like the Solar System \cite{Brax:2014c}, which in our metric sign convention (see next section) implies $\chi < 0$. In this regime, we shall also make use of the bounds imposed by the Cassini satellite \cite{Bertotti:2003rm} on the magnitude of fifth forces in the Solar System {and check that we satisfy the constraints provided by laboratory experiments}.

On cosmological scales, the dynamics of $\varphi$ become important, $\chi > 0$, which broadens the range of tests that can be used to constrain K-mouflage models. On the one hand, in the Jordan frame, where matter couples to gravity minimally, Newton's {\it constant} becomes locally time dependent due to the cosmological evolution of the scalar field \cite{Babichev:2011iz}. We shall show that this can be used to place constraints on the coupling strength $\beta$ (cf.~Eq.~(\ref{A-exp-def})) by using the results from the same Lunar Laser Ranging experiment mentioned above, which currently constrains the rate of change of the gravitational strength at the $10^{-12}$ yr$^{-1}$ level \cite{Williams:2004qba}. On the other hand, the formation of large-scale structure is also affected by the K-mouflage field \cite{Brax:2014a,Brax:2014b,Barreira:2014a}. Although the precision of cosmological data is not as good as that from local tests, one can still impose some constraints by requiring that the cosmological behavior should not differ too much from standard $\Lambda$-CDM.

In this paper, we first identify the requirements on the static ($\chi < 0$) and dynamical ($\chi > 0$) branches for K-mouflage models to comply with the current data, and then we attempt to design $K(\chi)$ functions that interpolate between these two branches in observationally and theoretically viable manners. A main result of this paper is that, although the small-scale constraints do limit significantly the functional forms allowed for $K(\chi)$, it is nevertheless possible to construct functions that exhibit percent-level modifications on the growth of structure relative to $\Lambda$-CDM. This is mainly because on larger scales the K-mouflage screening effect becomes less efficient, which can have an impact on the nonlinear matter power spectrum and halo mass functions, as we show using the results from semi-analytical models of structure formation. The size of the differences to $\Lambda$-CDM are within the ballpark of future observational missions such as Euclid \cite{2011arXiv1110.3193L} or LSST \cite{2012arXiv1211.0310L}, which makes these models predictive and therefore testable.

The paper is organized as follows. In Sec.~II, we introduce K-mouflage models and we summarize their main properties. In Sec.~III, we focus on the small-scale regime ($\chi < 0$), which applies to Solar System scales, and where we consider
the constraints from the anomalous perihelion of the Moon and the result from Cassini. We briefly recall the main aspects of the cosmological evolution and growth of large-scale structure in K-mouflage models in Sec.~IV. In Sec.~V, we analyze the connection between the time dependence of the gravitational strength and the cosmological evolution of $\varphi$. In Sec.~VI we summarize all  the constraints and we build models that satisfy them. Sec.~VII is devoted to the analysis of the cosmological dynamics of both the background and the matter density perturbations in the models constructed in Sec.~VI. Finally, we conclude in Sec.~VIII.
We discuss superluminality and causality issues in the appendix.

\section{The K-mouflage model}
\label{K-mouflage}

Following previous works \cite{Brax:2014a,Brax:2014b,Brax:2014c,Barreira:2014a},
we consider scalar field models where the action in the
Einstein frame has the form
\beqa
S & = & \int \dd^4 x \; \sqrt{-g} \left[ \frac{\MPl^2}{2} R + {\cal L}_{\varphi}(\varphi)
\right] \nonumber \\
&& + \int \dd^4 x \; \sqrt{-\tg} \, \tilde{\cal L}_{\rm m}(\tilde{\psi}^{(i)}_{\rm m},\tg_{\mu\nu}) ,
\label{S-def}
\eeqa
where
 $g$ is the determinant of the Einstein-frame metric tensor $g_{\mu\nu}$, and
$\tilde{\psi}^{(i)}_{\rm m}$ are various matter fields defined in the Jordan frame.
This also defines the Einstein-frame Newton's constant as $8\pi\cG=M_{\rm Pl}^{-2}$.
The K-mouflage scalar field $\varphi$ is explicitly coupled to matter through the
Jordan-frame metric $\tg_{\mu\nu}$, which is given by the conformal rescaling
\beq
\tg_{\mu\nu} = A^2(\varphi) \, g_{\mu\nu} ,
\label{g-Jordan-def}
\eeq
and $\tg$ is its determinant (here and throughout, quantities with a tilde are defined in the Jordan frame).
In this paper, we consider models with a non-standard kinetic term
\beq
{\cal L}_{\varphi}(\varphi) = \cM^4 \, K\left(\frac{X}{\cM^4}\right) \;\;\; \mbox{with} \;\;\;
X = - \frac{1}{2} \, \pl^{\mu}\varphi \, \pl_{\mu}\varphi .
\label{K-def}
\eeq
We use the signature $(-,+,+,+)$ for the metric.
Here, $\cM^4$ is an energy scale of the order of the current energy density of the Universe
(i.e., set by the cosmological constant), to recover the late-time accelerated expansion
of the Universe.

It is convenient to introduce the dimensionless kinetic energy $\chi$ by,
\beq
\chi = \frac{X}{\cM^4}  = - \frac{1}{2\cM^4} \; \pl^{\mu}\varphi\pl_{\mu}\varphi .
\label{chi-def}
\eeq
Then, as described in \cite{Brax:2014a}, the canonical behavior (i.e., $K \sim \chi \propto -(\pl\varphi)^2/2$),
with a cosmological constant $\rho_{\Lambda} = \cM^4$,  is recovered at late time in the weak-$\chi$ limit
if we have:
\beq
\chi \rightarrow 0 : \;\;\; K(\chi) \simeq -1 + \chi + ... ,
\label{K-chi=0}
\eeq
where the dots stand for higher-order terms. We shall impose this limit on all the models that we analyze.
The Klein-Gordon equation that governs the dynamics of the scalar field
$\varphi$ is obtained from the variation of the action (\ref{S-def}) with respect to
$\varphi$
\beq
\frac{1}{\sqrt{-g}} \pl_{\mu} \left[ \sqrt{-g} \; \pl^{\mu} \varphi \; K' \right] -
\frac{\dd\ln A}{\dd\varphi} \; \rho_E = 0 ,
\label{KG-1}
\eeq
which differs from the usual Klein-Gordon equation by having an extra term due to the coupling of the scalar field to matter, and where $\rho_E=- g^{\mu\nu}T_{\mu\nu}$ is the Einstein-frame matter density. A prime denotes partial differentiation  with respect to ~$\chi$, e.g.~$K'=\dd K/\dd\chi$.
For simplicity, we assume that $\beta=M_{\rm Pl} \dd \ln A/\dd\varphi$ is a constant, which implies
\beq
A(\varphi) = e^{\beta\varphi/M_{\rm Pl}} .
\label{A-exp-def}
\eeq

The normalization of the first two terms in Eq.(\ref{K-chi=0}) only defines the
normalization of the constant ${\cal M}^4$ and of the field $\varphi$
(except for its sign), and therefore it does not imply any loss of generality.
Similarly, the sign of $\beta$ in Eq.(\ref{A-exp-def}) can be absorbed in the sign of
$\varphi$, and therefore we can choose $\beta>0$ without any loss of generality.

For observationally interesting cases, we have $|A-1| \lesssim 0.1$ (see for instance \cite{Brax:2014a,Brax:2014b,Barreira:2014a} or the discussion in
Sec.~\ref{sec:Time-variation-of-G}).
Therefore, higher-order terms in $\ln A(\varphi)$ would only make small quantitative changes with a negligible impact on our conclusions. Next, we shall first analyze the solutions to the Klein-Gordon equation for spherically symmetric static configurations. Then we investigate the solutions in a cosmological setting, where the dynamics of $\varphi$ become important.

\section{Local Dynamics}
\label{sec:Local-Dynamics}

In this section we study the model predictions for Solar System scales, where we work in the Einstein frame. We assume the fields are static, which is a valid approximation since we consider time scales that are much shorter than the cosmological ones. In this case, $\bar{A}({\bar\varphi})$, associated with the slow running of the cosmological
background, is approximately constant and the Einstein and Jordan frames
are equivalent. \footnote{In Sec.~\ref{sec:Time-variation-of-G}, we shall relax the static approximation when we study the slow time variation of Newton's gravitational strength in the Jordan frame.} To simplify notations we drop the subscript ``E'' in this section and we denote $c$ as the
speed of light. One can split the scalar field as
\beq
\varphi_{\rm local}(\vr,t) = \bar{\varphi}(t) + \varphi(\vr) ,
\label{phi-local}
\eeq
where $\bar{\varphi}$ is the uniform value associated with the cosmological background and $\varphi(\vr)$ is the perturbed component on which  we focus in this section.
The cosmological background plays no role in this section, apart from setting the value of the coupling
factor $\beta \simeq \dd\ln\bar{A}/\dd\bar{\varphi}$.

\subsection{Static case}
\label{sec:static}

For a source with density $\rho$, the static Klein-Gordon equation becomes
(see \cite{Brax:2014c} for details)
\beq
\nabla_{\vr} \cdot (\nabla_{\vr} \varphi \; K' ) =  \frac{\beta\rho}{c^2 M_{\rm Pl}} ,
\label{KG-static-1}
\eeq
with $\chi= - c^2 (\nabla_{\vr}\varphi)^2/(2{\cal M}^4)$,
from which one can obtain a first-order algebraic equation for $\nabla_{\vr} \varphi$
\beq
\nabla_{\vr} \varphi \; K' = \frac{2\beta M_{\rm Pl}}{c^2}
( \nabla_{\vr} \Psi_{\rm N} + \nabla_{\vr} \times \vec\omega ),
\label{KG-omega-curl}
\eeq
where $\Psi_{\rm N}$ is the Newtonian potential, given by the usual Poisson equation
\beq
\nabla_{\vr}^2 \Psi_{\rm N} = 4\pi \cG \rho ,
\label{Poisson}
\eeq
and ${\vec\omega}$ is a divergence-free potential vector
(which must be determined along with $\varphi$).

The right-hand side of the Poisson equation also involves the fluctuations of the scalar field energy
density $\delta\rho_{\varphi}$, but as shown in Ref.~\cite{Brax:2014c}, this would only introduce negligible effects compared to the fluctuations of the matter density.

The spatial fluctuations of the coupling function $A(\varphi)$ can also be neglected in most cases
(see \cite{Brax:2014b,Brax:2014c} and Eq.~(\ref{Alocal-Abar}) below), except in the Euler equation
or the geodesic equation (\ref{5th-force}) below, which involve the gradient of $A(\varphi)$ that gives rise to the fifth force associated with the scalar
field gradient. In a similar fashion, we work in the weak gravitational field and non-relativistic limit,
so that the metric fluctuations $\Psi_{\rm N}$ only appear in the Euler equation
or the geodesic equation (\ref{5th-force}) below through the gradient $\nabla_{\vr}\Psi_{\rm N}$.
For the spherical configurations we consider here, $\vec\omega=0$, which allows one to analyze the dynamics of the system and the scalar force due to
$\varphi$ in a more straightforward manner.

The fifth force generated by the K-mouflage field  is given by \cite{Brax:2014c}
\beq
\vF_{\varphi} \equiv - \frac{\beta c^2}{M_{\rm Pl}} \nabla_{\vr}\varphi =
- \frac{2\beta^2}{K'} \nabla_{\vr} \Psi_{\rm N} .
\label{F-fifth-force}
\eeq
The K-mouflage screening mechanism relies on the fact that, in the nonlinear regime, i.e. deep inside the K-mouflage radius,
the factor $K'$ becomes large, which suppresses the fifth force relative to the Newtonian one, $\vF_{\rm N} = -\nabla_{\vr}\Psi_{\rm N}$ (note that
$|  \vF_{\varphi} | \sim | \vF_{\rm N} / K' |$). The implementation of the screening can be illustrated in a few steps (see also \cite{Brax:2014c}).
For a spherical matter distribution $\rho(r)$ with mass profile
\be
M(r)= \int_0^r \dd r' \; 4\pi r'^2 \, \rho (r'),
\ee
the Klein-Gordon equation (\ref{KG-omega-curl}) can be written as
\be
\frac{\dd\varphi}{\dd r} \; K' = \frac{\beta M(r)}{c^2 M_{\rm Pl} 4\pi r^2} , \;\; \mbox{with} \;\; \chi= - \frac{c^2}{2{\cal M}^4} \left(\frac{\dd\varphi}{\dd r}\right)^2 .
\label{KG-4}
\ee
We define the  ``K-mouflage screening radius'' $R_K$ by \cite{Brax:2014b,Brax:2014c}
\beq
R_K = \left( \frac{\beta M}{4\pi c M_{\rm Pl} {\cal M}^2} \right)^{1/2} ,
\label{RK-def}
\eeq
where $M=M(R)$ is the total mass of the object of radius $R$.
Then, by introducing the rescaled dimensionless variables $x=r/R_K$, $m(x)= M(r)/M$ and $\phi(x)= \varphi(r)/\varphi_K$, with
\beq
\varphi_K = {\cal M}^2 R_K /c ,
\label{phiK-def}
\eeq
the integrated Klein-Gordon equation (\ref{KG-4}) becomes
\beq
\frac{\dd\phi}{\dd x} \, K' = \frac{m(x)}{x^2}
, \;\; \mbox{with} \;\; \chi= - \frac{1}{2} \left( \frac{\dd\phi}{\dd x} \right)^2 ,
\label{KG-5}
\eeq
which can also be written as
\beq
\sqrt{-2\chi} K'(\chi) = \frac{m(x)}{x^2} .
\label{chi-Kp}
\eeq
A unique solution of the Klein-Gordon equation is always guaranteed when
$\sqrt{-2\chi} K'(\chi)$ is a monotonic decreasing function over $\chi<0$,
which grows up to $+\infty$ as $\chi \rightarrow -\infty$. This is assumed in the following
\footnote{See \cite{Brax:2014c} for a more detailed study of this case, and the relaxation of the scalar field
to its equilibrium state, as well as a discussion of badly behaved cases where there are no
well-defined static profiles, i.e. when $\sqrt{-2\chi} K'(\chi)$ is not a monotonic function that decreases from $+\infty$ to zero, as $\chi$ varies from $-\infty$ to zero.}
and it implies $K'+2\chi K'' >0$.
A test particle outside the matter distribution evolves according to the non-relativistic equation of motion
\be
\frac{\dd^2 \vr}{\dd t^2}= - \nabla_{\vr} \Psi_{\rm N}
- \frac{\beta c^2}{M_{\rm Pl}} \nabla_{\vr} \varphi ,
\label{5th-force}
\ee
which becomes the same as in standard gravity, provided one interprets the equation with a total potential that is the sum of the fifth force one $\delta\Psi$
\beq
\delta\Psi= \frac{\beta c^2}{M_{\rm Pl}} \varphi,
\label{dPsi-def}
\eeq
with the Newtonian potential $\Psi_{\rm N}$. For a spherical body we can consider radial trajectories and the scalar field gradient is
given by Eq.~(\ref{KG-5})
\beq
\frac{\dd\varphi}{\dd r} = \frac{\varphi_K}{R_K} \; \frac{m(x)}{x^2 K'} .
\eeq
Outside the spherical body we have $m(x)=1$ and we obtain
the equation of motion
\beq
\frac{\dd^2 r}{\dd t^2} = - \frac{\cG M}{r^2} \; \left( 1+ \frac{2\beta^2}{K'(\chi(r))} \right) .
\label{radial-1}
\eeq
This equation shows that the standard gravitational law acquires extra terms proportional to $\beta^2/K^\prime$. This can be captured by defining an effective gravitational strength
\beq
\cG^{\rm eff}(r) = \left( 1+ \frac{2\beta^2}{K'(\chi(r))} \right) \; \cG
\label{Geff-1}
\eeq
which depends on the distance from the central object. As the test particle gets deeper inside the K-mouflage radius, then $m/x^2 \gg 1$ and $K'(\chi(r))$ becomes larger. If $K'(\chi(r))$ becomes large enough, then the correction term in Eq.~(\ref{Geff-1}) becomes negligible and Newton's law is retrieved. Next, we investigate the conditions for this correction to be small enough to evade the constraints on the anomalous perihelion of the Moon set by Lunar Laser Ranging experiments \cite{Williams:2012a}.

\subsection{Perihelion constraints}
\label{sec:Perihelion}

\subsubsection{Constraint on the kinetic function}
\label{sec:peri-K}

The explicit dependence on $r$ of the correction to Newton's law implies that orbits are perturbed and in particular that there is an anomalous perihelion for objects like the Moon orbiting around the Earth.
The perihelion angular advance is given by
\beq
\delta\theta = \pi r \frac{\dd}{\dd r} \left[ r^2 \frac{\dd}{\dd r} \left( \frac{\varepsilon}{r} \right) \right] ,
\label{perihelion-def}
\eeq
where $\varepsilon$ is the ratio between the fifth-force and Newtonian potentials
\beq
\varepsilon = \frac{\delta\Psi}{\Psi_{\rm N}}
= \frac{\beta c^2 \varphi}{M_{\rm Pl} \Psi_{\rm N}}
= - \frac{\beta c^2 r \varphi}{M_{\rm Pl}{\cal G} M} ,
\label{epsilon-def}
\eeq
and where we have used $\Psi_{\rm N}=-{\cal G}M/r$.
This gives explicitly
\beq
\frac{\varepsilon}{r} = -  \frac{\beta c^2 \varphi}{M_{\rm Pl}{\cal G} M} ,
\label{eps-r}
\eeq
and, using Eq.~(\ref{KG-4}), one gets the variation
\beq
\frac{\dd}{\dd r} \left( \frac{\varepsilon}{r} \right)  = - \frac{2\beta^2}{K' r^2} .
\eeq
This implies that the anomalous perihelion is given by
\beq
\delta\theta = \pi r \frac{\dd}{\dd r} \left[ \frac{-2\beta^2}{K'} \right]
= 2\pi \frac{\beta^2}{K'^2} x K'' \frac{\dd\chi}{\dd x} ,
\label{dtheta-Ks}
\eeq
where, we recall, $x=r/R_K$.
Using the fact that, outside the spherical source the Klein-Gordon equation (\ref{chi-Kp}) yields
\beq
\sqrt{-2\chi} K'(\chi) = \frac{1}{x^2} =  \left( \frac{R_K}{r} \right)^2 ,
\label{chi-Kp-1}
\eeq
then, its derivative with respect to $\chi$ gives
the spatial variation of  $\chi$ as
\beq
\frac{\dd x}{\dd \chi} = - \frac{x}{4\chi} \frac{c_s^2}{c^2} ,
\eeq
where we have defined the speed of scalar spherical waves around the massive object
\cite{Brax:2014c}
\beq
c_s^2 = \frac{K'+2\chi K''}{K'} c^2 > c^ 2 .
\label{cs-def}
\eeq
This yields our final result for the anomalous perihelion advance
\beq
\delta\theta = -8\pi \frac{\beta^2}{K'} \frac{\chi K''}{K'} \frac{c^2}{c_s^2} ,
\label{theta-cs}
\eeq
as expressed solely in terms of $\chi$ and of the Lagrangian function $K$ and its derivatives. This expression contains the factor $\beta^2/K'$, which, as we have seen in Eq.~(\ref{Geff-1}) controls the amplitude of the fifth force. The Cassini satellite results constrain this amplitude in the Solar System in such a way that $\frac{2\beta^2}{K'}<10^{-5}$ \cite{Bertotti:2003rm}. On the other hand, the Lunar Ranging experiment implies for the Earth-Moon system that
$\vert\delta \theta\vert \le 2\times10^{-11}$. Thus, the Lunar Ranging constraint is much stronger than the Cassini bound and will prove
to be the main source of constraints on the form of $K(\chi)$ on the static branch $\chi<0$ (cf.~Sec.~\ref{sec:Constraining the models} below). The only sensible way of complying with this bound is to reduce $\chi K''/K'$ in the static case, which also gives $c_s \simeq c$. This can be achieved by suppressing the contribution of the nonlinear terms in $K(\chi)$ when $\chi < 0$.

\subsubsection{Constraint on the running of the coupling $\beta$}
\label{sec:peri-beta}

Throughout this paper, we focus on the case of a constant coupling strength $\beta$,
which corresponds to the exponential coupling function (\ref{A-exp-def}).
However, we may also consider more general coupling functions where $\beta$
would now depend on time and space through the variations of the scalar field $\varphi$.
This would not change our results for the kinetic function $K(\chi)$ and the typical
amplitude of the coupling $\beta$, because $|\varphi/M_{\rm Pl}|$ (which goes to zero
at high redshift) does not grow beyond $0.5$ for  observationally interesting models,
as seen in Eqs.(\ref{phibar-estimate}) and (\ref{Alocal-Abar}) below.
Hence the variations of $\beta$ are small in realistic models.

On the other hand, we can investigate whether the very small bound on the anomalous
perihelion, $\vert\delta \theta\vert \le 2\times10^{-11}$, provides interesting constraints
on the possible amount of running of $\beta$.
For a general coupling function $A(\varphi)$ the fifth-force potential reads as
$\delta\Psi= c^2 \ln(A/\bar{A})$ and the perihelion advance (\ref{perihelion-def})
gives
\beq
\delta\theta = \pi r \frac{\dd}{\dd r} \left[ - \frac{r^2 c^2}{{\cal G} M} \frac{\beta}{M_{\rm Pl}}
\frac{\dd\varphi}{\dd r} \right ] ,
\eeq
where we used again $\Psi_{\rm N}=-{\cal G}M/r$.
If $\beta$ depends on space, the Klein-Gordon equation (\ref{KG-static-1}) cannot
be exactly integrated as in Eq.(\ref{KG-4}). However, the fluctuations of $\beta$ can be
neglected at first order, and hence we can still approximate $\dd\varphi/\dd r$ by
Eq.(\ref{KG-4}).
This yields
\beq
\delta\theta = \pi r \frac{\dd}{\dd r} \left[ - \frac{2\beta^2}{K'} \right]
= - 4 \pi \frac{\beta}{K'} r \frac{\dd\beta}{\dd r} + 2\pi \frac{\beta^2}{K'^2} r \frac{\dd K'}{\dd r} .
\eeq
The second term is the one that was already obtained in Eq.(\ref{dtheta-Ks}) for
constant $\beta$. Focusing on the first term and using again Eq.(\ref{KG-4}), we obtain
the contribution to the anomalous perihelion due to the spatial variation of $\beta$ as
\beq
\delta\theta = 8\pi \frac{\Psi_{\rm N}}{c^2} \frac{\beta^2 \beta'}{K'^2} .
\label{dtheta-beta}
\eeq
Here we defined
\beq
\beta = M_{\rm Pl} \frac{\dd\ln A}{\dd\varphi} , \;\;\;
\beta' = M_{\rm Pl} \frac{\dd\beta}{\dd\varphi}
= M_{\rm Pl}^2 \frac{\dd^2\ln A}{\dd\varphi^2} ,
\label{beta-p-def}
\eeq
the dimensionless derivative of the coupling with respect to $\varphi/M_{\rm Pl}$.
We shall see in Sec.~\ref{sec:running} that, because of the small prefactor
$\Psi_{\rm N}/c^2$, the bound on the perihelion advance only gives a very loose
bound on the derivative $\beta'$, which does not provide useful
information on the coupling function $A(\varphi)$.
This is consistent with the fact that fluctuations of the scalar field $\varphi$ and of
$A$ can be neglected in most places, except as the source of the fifth
force that explicitly involves the gradient of $A$.

\subsection{Laboratory tests}
\label{sec:laboratory}

{Measurements of the orbits of planets in the Solar System constrain the deviations from General
Relativity, through Eq.(\ref{Geff-1}) or Eq.(\ref{theta-cs}).
In addition, laboratory experiments, such as the ones using atom interferometry and measuring the acceleration
induced by a test mass of a few grams over distances of a few centimeters, also constrain the amplitude of
the fifth force in Eq.~(\ref{Geff-1}) to a $10^{-4}$ accuracy \cite{Sorrentino:2013uza,2008PhRvL.100e0801L}.
Thus laboratory experiments place constraints on $2 \beta^2/K'$ in the static case, but in an even more
non-linear regime than the Cassini spacecraft or the Lunar Ranging experiment.
Indeed, laboratories on the surface of the Earth are further inside the K-mouflage radius of the Earth than
the Moon.}

\section{Cosmology}
\label{sec:screening}

Before analyzing the constraints on K-mouflage models obtained from regimes where the dynamics of $\varphi$ are important ($\chi > 0$), we first briefly recap the equations relevant for the cosmological evolution of the background and of linear perturbations. For further details, we refer the reader to \cite{Brax:2014a} for a study of the background expansion history, to \cite{Brax:2014b} for a study of large-scale structure formation and to \cite{Barreira:2014a} for a study of the model predictions for the Cosmic Microwave Background (CMB) temperature and lensing potential power spectra.

\subsection{Background}
\label{sec:Background}

Considering only spatially-flat universes, the Einstein equations lead to the usual Friedmann equations \cite{Brax:2014a},
\beqa
3 \MPl^2 H^2 & = & \bar{\rho}_E + \bar{\rho}_{\varphi} ,
\label{Friedmann-1} \\
-2 \MPl^2 \dot{H} & = & \bar{\rho}_E + \bar{\rho}_{\varphi} + \bar{p}_{\varphi}
\label{Friedmann-2}
\eeqa
where $\rho_E$, $\rho_{\varphi}$ and $p_{\varphi}$, are, respectively, the matter and scalar
field energy densities and pressure in the Einstein frame:
\beq
\bar{\rho}_{\varphi} = - \cM^4 \bar{K} + \dot{\bar\varphi}^2 \, \bar{K}'  , \;\;\;
\bar{p}_{\varphi} = \cM^4 \bar{K} .
\label{rho-phi-def}
\eeq
The overbar denotes uniform background quantities,
and the dimensionless field $\chi$ (Eq.~(\ref{chi-def})) simplifies as
\beq
\bar{\chi} = \frac{\dot{\bar\varphi}^2}{2\cM^4} .
\label{chi-bar-def}
\eeq
 It is convenient to introduce the rescaled matter
density $\rho$,
\beq
\rho = A^{-1} \rho_E ,
\eeq
which satisfies the standard conservation equation.
The Klein-Gordon equation (\ref{KG-1}) is now given by
\beq
\frac{\dd}{\dd t} \left( a^3 \dot{\bar\varphi} \bar{K}' \right) =
- \frac{\dd \bar{A}}{\dd\bar\varphi} \, \bar{\rho} \, a^3.
\label{KG-2}
\eeq
We can also define an effective scalar field energy
density
\beq
\rho_{\varphi}^{\rm eff} = \rho_{\varphi} + [ A(\varphi)-1] \rho ,
\label{rho-phi-eff-def}
\eeq
which satisfies the standard conservation equation (the pressure $p_{\varphi}$
is not modified \cite{Brax:2014a})
\beq
\dot{\bar{\rho}}_{\varphi}^{\rm eff} = - 3 H (\bar{\rho}_{\varphi}^{\rm eff}
+ \bar{p}_{\varphi} ) .
\label{conserv-1}
\eeq
Under these definitions, the Friedmann equations (\ref{Friedmann-1})-(\ref{Friedmann-2}) become
\beqa
3 \MPl^2 H^2 & = & \bar{\rho} + \bar{\rho}_{\varphi}^{\rm eff} ,
\label{Friedmann-3} \\
-2 \MPl^2 \dot{H} & = & \bar{\rho} + \bar{\rho}_{\varphi}^{\rm eff} + \bar{p}_{\varphi}.
\label{Friedmann-4}
\eeqa
We define also the time-dependent cosmological parameters
\beq
\Omega_{\rm m} = \frac{\bar\rho}{\bar\rho+\bar{\rho}_{\varphi}^{\rm eff}} , \;\;
\Omega_{\varphi}^{\rm eff} = \frac{\bar{\rho}_{\varphi}^{\rm eff}}
{\bar\rho+\bar{\rho}_{\varphi}^{\rm eff}} , \;\;
w_{\varphi}^{\rm eff} = \frac{\bar{p}_{\varphi}}{\bar{\rho}_{\varphi}^{\rm eff}} .
\label{w-def}
\eeq

At early times, we have $\bar\varphi \rightarrow 0$ and $A(\varphi)$ is normalized
by $A(0)=1$ \cite{Brax:2014a}. For observationally interesting models, we have $A \sim 1$ ($|A-1| \lesssim 0.1$) until today \cite{Brax:2014a,Barreira:2014a}
(see also the discussion on the constraints from Big Bang Nucleosynthesis (BBN) below). From Eq.~(\ref{KG-2}), one can then write
\beq
\dot{\bar\varphi} \sim -  \frac{\beta \bar\rho t}{M_{\rm Pl} \bar K'} , \;\;\;
\frac{\beta\bar\varphi}{M_{\rm Pl}} \sim -\frac{\beta^2}{\bar K'} .
\label{phibar-estimate}
\eeq

\subsection{Linear Perturbations}
\label{sec:Linear-Perturbations}

On large scales, the evolution modes, $D_{\pm}(\eta)$, of small linear density fluctuations satisfy the equation \cite{Brax:2014b}
\beq
\frac{\dd^2 D}{\dd\eta^2} + \left(
\frac{1-3w_{\varphi}^{\rm eff}\Omega_{\varphi}^{\rm eff}}{2} + \epsilon_2 \right)
\frac{\dd D}{\dd\eta} - \frac{3}{2} \Om (1+\epsilon_1) D = 0,
\label{D-linear}
\eeq
where $\eta=\ln a$ is the number of $e$-foldings. The functions $\epsilon_1$ and $\epsilon_2$ are time dependent only and they are given by
\beq
\left| \epsilon_1 \right| = \left| \bar{A}-1+\frac{2\bar{A}\beta^2}{\bar{K}'} \right|
\sim \left| \frac{\beta^2}{\bar{K}'}  \right| ,
\label{eps1-1}
\eeq
and
\beq
\epsilon_2= \frac{\dd\ln \bar{A}}{\dd\eta} =
\frac{\beta}{M_{\rm Pl}} \frac{\dd\bar\varphi}{\dd\eta}
\sim - \frac{\beta^2}{\bar{K}'} .
\label{eps2-1}
\eeq
In Eq.~(\ref{eps1-1}) the sign of $\epsilon_1$ cannot be determined {\it a priori} because the terms $(\bar{A} -1)
\simeq \beta\bar\varphi/M_{\rm Pl}$ and $2\bar{A}\beta^2/\bar{K}'$ are typically of opposite
signs and of the same order.
In Eq.~(\ref{D-linear}), $\epsilon_1$ mimics the effects of a modified Newton's gravitational strength on the linear growth of structure, while $\epsilon_2$
appears as a friction term. Both these terms are of order $\beta^2/\bar{K}'$, just like the case of $\bar{\varphi}$ in Eq.~(\ref{phibar-estimate}).

The reader might note that whereas in the cosmological linear equation
(\ref{D-linear}) Newton's gravitational strength appears to be amplified by a factor
$(1+\epsilon_1)$, in the static case (\ref{Geff-1}) we found a factor $(1+2\beta^2/K')$
and in Eq.(\ref{G-drift}) below we have a slow drift given by $A^2$.
These differences come from the fact that they apply to different regimes,
which also involve different sets of approximations.
This explicitly shows that modifications of gravity, as defined from the Lagrangian of
the theory, can have subtle effects that are not captured by a unique rescaling of
Newton's gravitational strength.

More precisely, the linear evolution equation (\ref{D-linear}) applies to the
cosmological density field in the Einstein frame, where the energy-momentum
tensor is not conserved because of the time-dependent conformal mapping
(\ref{g-Jordan-def}) to the Jordan frame that defines the matter Lagrangian.
This slow drift between the Einstein and Jordan frames gives rise to the factors
$\bar{A}$ in Eqs.(\ref{eps1-1}) and (\ref{eps2-1}).
In contrast, the static equation (\ref{Geff-1}) applies to small-scale systems
over time scales that are short as compared with the Hubble time, so that
the cosmological variation of $\bar{A}$ and the expansion of the Universe
can be neglected. Then, non-conservation terms of the form
$\rho \dd\ln \bar{A}/\dd t$ can be neglected and there is no difference between
the Einstein- and Jordan-frame density fields [thus, with $\bar{A} \equiv 1$
Eq.(\ref{eps2-1}) gives $\epsilon_2=0$ and
Eq.(\ref{eps1-1}) gives back $\epsilon_1=1+2\beta^2/\bar{K'}$, as in
Eq.(\ref{Geff-1}), except that $\bar{K}'$ is the cosmological background value
of the kinetic function whereas in Eq.(\ref{Geff-1}) it is the local perturbed value
associated with the compact object].
Finally, Eq.(\ref{G-drift}) gives the cosmological drift of Newton's coupling in the
Jordan frame, as opposed to the Einstein-frame evolution equation (\ref{D-linear})
(where by definition Newton's coupling is indeed constant) and to the
static equation (\ref{Geff-1}) (where cosmological drifts are neglected as compared
with the system time scale, such as the orbital period of the planet).
Thus, these differences come from the fact that we consider different
time and length scales, and different frames.

\section{Time variation of $\cG$}
\label{sec:Time-variation-of-G}

In Sec.~\ref{sec:Local-Dynamics}, we have neglected the effects of the cosmological evolution on the predictions for the Solar System, since we considered time scales that are much shorter than the Hubble scale.
However, in K-mouflage models, the conformal mapping of Eq.~(\ref{g-Jordan-def}) implies that in the Jordan frame, where matter couples minimally,
Newton's gravitational strength, $\tilde{\cG}$, should be time varying. This can be understood as follows. Let us consider the conformal transformation from the Einstein to the Jordan frame. Since, the Ricci scalar and $\sqrt{-g}$ transform, respectively, as $R= A^2 \tilde{R} + ...$, (where the dots stand for additional terms associated with derivatives of $A$), and $\sqrt{-g}=A^{-4} \sqrt{-\tilde{g}}$, then one has $\tilde{\cG} = A^2 \cG = A^2/8\pi M_{\rm Pl}^2$. That is, $\tilde{\cG}$ becomes time-varying due to the background evolution of $\bar{A}(\bar{\varphi})$. This time variation of Newton's gravitational strength in the Jordan frame can be constrained in two ways: (a) through the comparison between the local value of $\tilde{\cG}$
and that at the time of BBN, and (b) through the impact on the trajectories of planets and moons in the Solar System.

Starting with the BBN case, a value of Newton's constant which would be different during BBN compared to the one inferred from
local measurements would be tantamount to a change in the Hubble rate and therefore would lead to a
discrepancy in the formation of the elements. Such a change cannot exceed about ten percent
\cite{Uzan:2010pm,Agnetta:2007um}. In the Jordan
frame we have
\beq
\tilde{\cG} =A^2(\varphi) \cG \approx \left( 1 + \frac{2 \beta \varphi}{M_{\rm Pl}} \right) \cG ,
\label{G-drift}
\eeq
which implies the bound
\beq
\frac{\beta}{M_{\rm Pl}} \left| {\bar\varphi}_{\rm BBN} - \varphi_{\rm local} \right| \lesssim 0.05 .
\label{BBN-1}
\eeq
At the time of BBN, we have ${\bar\varphi}_{\rm BBN} = \bar{\varphi}(z\sim10^{10}) \simeq 0$. The local value of the scalar field is given by Eq.(\ref{phi-local}). In Sec.~\ref{sec:Local-Dynamics}, we neglected the contribution from the background part, $\bar{\varphi}$, since we were interested only in the additional $r$-dependence in the force law. Here, however, we should take it into account, and from Eq.~(\ref{phibar-estimate}) we have
$\beta\bar\varphi/M_{\rm Pl} \sim \beta^2/\bar{K}' \sim \beta^2$ (because $\bar{K}' \simeq 1$ at low redshifts). From Eq.~(\ref{F-fifth-force}), we have that the perturbed part of the scalar field, $\varphi(r)$, is of order $\beta\varphi(r)/M_{\rm Pl} \sim \beta^2 \Psi_{\rm N}/K' c^2 \ll \beta^2$, because $K' \gg 1$
in the Solar System (see Sec.~\ref{sec:Combined-Constraints} below) and
$\Psi_{\rm N}/c^2 \ll1$ (in the Solar System we have $\Psi_{\rm N}/c^2 \sim 10^{-6}$). Altogether we have
\beq
| \varphi(r) | \ll | \bar{\varphi}(t) | \;\;\; \mbox{and} \;\;\; A(\varphi_{\rm local}) \simeq \bar{A} ,
\label{Alocal-Abar}
\eeq
and Eq.~(\ref{BBN-1}) implies
\beq
\beta^2 \lesssim 0.05 .
\label{BBN-2}
\eeq

A second type of constraint on the rate of change of the gravitational strength comes from the change with time of the trajectories of planets and moons. This has been monitored by the Lunar Laser Ranging experiments for the Earth-Moon system \cite{Williams:2004qba}. Recalling that $\tilde{\cG} \simeq \bar{A}^2 \cG$, then its rate of change is given by
\beq
\frac{\dd\ln\tilde{\cG}}{\dd\tilde{t}} = \frac{2\epsilon_2}{1+\epsilon_2} \tilde{H}
\approx 2 \epsilon_2 \tilde{H} ,
\label{dG-1}
\eeq
where $\tilde H$ and $\tilde{t}$ are the Hubble rate and the time in the Jordan frame
(which are related to the Einstein-frame ones by $\tilde{H}=H(1+\epsilon_2)/\bar{A}$ and
$\dd\tilde{t}=\bar{A}\dd t$), and we have used Eq.~(\ref{eps2-1}).
Hence we find a direct link between the behavior of cosmological density perturbations and the Lunar Ranging constraint.
The current bound gives that
$\vert \dd\ln\tilde{\cG}/\dd\tilde{t} \vert_{\rm now} \lesssim 10^{-12}$ yr$^{-1}$ \cite{Williams:2004qba}.
In particular, taking as a reference value $h=0.67$ (although this value is not critical for the conclusions) we find that the constraint on $\epsilon_2$ reads
\be
\left| \epsilon_2 \right|_{\rm now} \lesssim 7.3\times10^{-3}.
\ee
From Eq.~(\ref{eps2-1}) we can see that this gives a constraint on the ratio
$\beta^2/\bar{K}'$ today. In fact, this is a strong constraint on the coupling $\beta$
of the K-mouflage field to matter, {which is independent} of the details of the kinetic
function $K(\chi)$. At late times, $\bar\chi$ goes to zero, and we typically
have $\bar\chi\ll 1$ today, as well as, $K' \simeq 1$, from Eq.~(\ref{K-chi=0}).
Therefore, we find
\be
\beta\lesssim 10^{-1} ,
\label{Lunar-1}
\ee
which is a tighter bound than the BBN constraint of Eq.~(\ref{BBN-2}).

In the Einstein frame, the variation of $\cG$ with time is replaced by a variation of the masses of fundamental particles as $ m_\psi= \bar A m_\psi^0$ where $m_\psi^0$ is the mass in the Jordan frame. In particular, we can see
that the ratio $m_\psi/M_{\rm Pl}$ of fundamental particle masses over the Planck scale is frame-invariant. For bound states such as the protons, and as long as the QCD phase transition can be modeled in the Jordan frame
where  no coupling of gluons to the scalar field are present, the masses in the Einstein frame are still proportional to $\bar A$. This guarantees that the proton-to-electron mass ratio $m_p/m_e$ is independent of time and no constraint
on K-mouflage can be drawn from the bounds on the variations of $m_p/m_e$ from quasar absorption lines.

From Eqs.~(\ref{eps1-1}) and (\ref{eps2-1}), ones notes that $\epsilon_1$ and $\epsilon_2$ are of the same order. Consequently, the bound that Lunar Ranging tests place on $\epsilon_2$, translates also into similar bounds to $\epsilon_1$, which highlights an interesting connection between Solar System constraints and the growth of structure on cosmological scales. In particular, from the above bounds one expects deviations from standard $\Lambda$-CDM to be of the order of a few percent \cite{Brax:2014b, Barreira:2014a}. On the other hand, the Lunar Ranging test does not constrain the past behavior of the model, when $\bar{\chi}$ deviates from zero. Therefore one must still investigate the cosmological impact of the functional form of $K(\chi)$ when $\chi \gtrsim 1$. This shall be done in part of the discussion of the next section.

\section{Constraining the models}
\label{sec:Constraining the models}

In this section, we summarize the constraints on K-mouflage models discussed above and also other stability conditions presented previously in the literature \cite{Brax:2014a, Brax:2014b, Brax:2014c} We also build models that satisfy them.

\subsection{Combined Constraints}
\label{sec:Combined-Constraints}

The cosmological regime corresponds to $\chi>0$, with $\chi\rightarrow+\infty$ at early
times and $\chi\rightarrow 0$ at late times. To avoid ghosts we must have $K'>0$
and $K'+2\chi K''>0$ \cite{Brax:2014a}, as fluctuations $\delta\varphi$ around the
cosmological background $\bar{\varphi}$ propagate with a speed $\bar{c}_s$
given by
\beq
\bar{c}_s^2 = \frac{\bar{K}'}{\bar{K}'+2\bar{\chi} \bar{K}''} c^2 < c^ 2 .
\label{bar-cs}
\eeq
This is formally the inverse of Eq.(\ref{cs-def}), but in Eq.(\ref{bar-cs})
$\chi=\bar{\chi}>0$ is the homogeneous time-dependent cosmological background,
whereas in Eq.(\ref{bar-cs}) $\chi<0$ is the small-scale static solution.
The function $\sqrt{\chi}K'(\chi)$ must also increase monotonically up to $+\infty$,
so that we have a well-defined cosmological behavior up to high redshifts
\cite{Brax:2014a,Brax:2014c}, which again implies $K'+2\chi K''>0$.
Requiring that the dark energy component becomes negligible with respect to the
matter density at early times implies that $K(\chi)$ grows faster than $\chi$
(e.g., as a power law $\chi^m$ with $m>1$, see Ref.~\cite{Brax:2014a}).
The marginal case, where $\bar\rho_{\varphi}$ grows as $1/t^2$ in the early matter era
(like the matter density) but is a small fraction of the matter density, corresponds to a
constant $\bar{K}'$ with $\bar{K}' \gg 1$.
Moreover, one can also impose that the growth of large-scale structures should not differ too much from $\Lambda$-CDM. A reasonable maximum deviation can be placed at the 
few-percent level today, which leads to the bound $\beta^2/\bar{K}' \lesssim 10^{-2}$, from Eqs.~(\ref{eps1-1})-(\ref{eps2-1}). This requires $\beta^2 \lesssim 10^{-2}$ because $\bar{K}' \simeq 1$ today.
This bound also ensures that the BBN constraint, Eq.~(\ref{BBN-2}), is satisfied. Therefore, the cosmological constraints are:
\[
\mbox{cosmological constraints:} \;\;\; \mbox{semi-axis} \;\; \chi > 0 ,
\]
\beq
K'>0 ,  \;\;  K'+2\chi K'' > 0 , 
\label{cosmo-constraints-Kp-Ks}
\eeq
\beq
\sqrt{\chi} K'(\chi) \rightarrow +\infty \;\; \mbox{for} \;\; \chi \rightarrow +\infty ,
\label{cosmo-constraints-Wp}
\eeq
\beq
K' \gg 1 \;\; \mbox{for} \;\; \chi \gg 1 , \;\; \beta \lesssim 0.1 .
\label{cosmo-constraints}
\eeq

The small-scale static regime corresponds to $\chi < 0$. 
To avoid singular behaviors and to ensure well-defined solutions for any matter density
profiles, we must have $K' >0$ and $\sqrt{-\chi} K'(\chi)$ monotonic and unbounded over
$\chi<0$ \cite{Brax:2014c} [recall also the discussion about Eq.~(\ref{chi-Kp})].
This latter condition also implies $K'+2\chi K''>0$ and ensures that the $c_s^2>0$,
where $c_s$ given in Eq.(\ref{cs-def}) is the propagation speed around a static
background.
Thus, we have the very general small-scale conditions:
\[
\mbox{small-scale constraints:} \;\;\; \mbox{semi-axis} \;\; \chi < 0 ,
\]
\beq
K'>0 ,  \;\;  K'+2\chi K'' > 0 , 
\label{static-constraints-Kp-Ks}
\eeq
\beq
\sqrt{-\chi} K'(\chi) \rightarrow +\infty \;\; \mbox{for} \;\; \chi \rightarrow -\infty .
\label{static-constraints}
\eeq
Note that instead of $\sqrt{-\chi} K'(\chi) \rightarrow +\infty$ at infinity it may be sufficient to go to infinity
at a finite negative value of $\chi$.
In addition, the Solar System dynamics corresponds to $\chi < -\chi_*$, as we are far 
in the nonlinear regime, where we denote by $(-\chi_*)$ the transition between the 
linear regime, where $K'\simeq 1$ and the kinetic function is governed by the first two terms of the expansion (\ref{K-chi=0}), and the nonlinear regime where $K' \gg 1$:
\beqa
\lefteqn{ \mbox{small-scale highly nonlinear regime:}  \;\;\; \chi < -\chi_*  , }  \nonumber \\
&& \;\;\; \mbox{with} -\chi_*<0 \;\; \mbox{and} \;\; K'(-\chi_*) \gg 1.
\label{chi*-def}
\eeqa

The Cassini bound on the amplitude of fifth forces in the Solar System implies that the scalar field should be sufficiently screened locally. From Eq.~(\ref{Geff-1})
this requires that
\beq
\mbox{Solar System screening:} \;\;\; \frac{\beta^2}{K'} \le 10^{-5}  \;\; \mbox{for}
\;\; \chi \sim \chi_{\rm s.s.}
\label{screening-constraints-1}
\eeq
where $K'$ must be evaluated at values $\chi_{\rm s.s.}$ that correspond to the
Solar System regime, that is, at distances of order one AU from the Sun.
This is well within the K-mouflage radius of the Sun itself, which means that
$\chi_{\rm s.s.} < -\chi_*$ and large values of $K'$ in this regime are consistent with
the requirement $K'(0)=1$.
More precisely, using that $\cM^4 \sim \bar\rho_{\rm de 0}$ is roughly the mean dark
energy density today, then we obtain from Eq.~(\ref{RK-def}) that the K-mouflage radius of
an object of mass $M$ is given by
\beq
R_K(M) \simeq \sqrt{\frac{\beta M}{M_{\odot}}} \; 3470 \; \mbox{AU} .
\label{RK-M}
\eeq
Thus, for $\beta \sim 0.1$, the K-mouflage radius of the Sun is
$R_K(M_{\odot}) \sim 1097$ AU, which is much larger than the size of the orbits of all Solar System major
bodies (Neptune and Pluto being at about $30$ and $40$ AU).
Moreover, the integrated Klein-Gordon equation (\ref{chi-Kp-1}) gives for the Solar System
regime:
\beq
\sqrt{-\chi_{\rm s.s.}} K_{\rm s.s.}' \sim 10^6 .
\label{chi-Kp-ss}
\eeq
In practice, the constraint (\ref{screening-constraints-1}) means that we require
\beq
K' \gg 1 \;\; \mbox{for} \;\;  \chi \ll -\chi_* ,
\label{screening-constraints-2}
\eeq
which automatically ensures that the general conditions (\ref{static-constraints}) are
satisfied.
Alternatively, one could also have very small values of the coupling parameter $\beta$,
but this would yield a cosmology virtually indistinguishable from $\Lambda$-CDM, rendering the K-mouflage scenario less interesting. In particular, if we assume $\beta \sim 0.1$, so that deviations from the $\Lambda$-CDM cosmology are not completely negligible, we require $K' > 10^3$ for $\chi \ll -\chi_*$.

From the constraints on the anomalous perihelion of the Moon, Eq.~(\ref{theta-cs}), we have
\beqa
\mbox{Moon perihelion:}&&  \;\;\; \frac{\beta^2}{K'}
\frac{|\chi K''|}{K'+2\chi K''}  \leq 8\times10^{-13} \nonumber \\
&& \;\; \mbox{for} \;\; \chi \sim \chi_{\rm m.e.} ,
\label{perihelion-constraint}
\eeqa
where $\chi_{\rm m.e.}$ corresponds to the Earth-Moon system.
The K-mouflage radius of the Earth is about 2 AU, and the distance between the Earth and
the Moon is $d \simeq 2.6 \times 10^{-3}$ AU, which gives
$[R_K(M_{\oplus})/d]^2 \sim 0.6 \times 10^{6}$.
Therefore, the Sun and the Earth have about the same impact on the scalar field
configuration at the location of the Moon. In practice, this means that the value of $\chi$ associated with the perihelion
constraint is roughly the same as that associated with the
Cassini bound (\ref{screening-constraints-1}):
\beq
\chi_{\rm m.e.} \sim \chi_{\rm s.s.} .
\label{chi-me-ss}
\eeq
In other words, the constraints of Eqs.~(\ref{screening-constraints-1}) and (\ref{perihelion-constraint}) both apply to the regime set by Eq.~(\ref{chi-Kp-ss}).

{Laboratory experiments give a constraint on the Newtonian force of order $10^{-4}$, which means
from Eq.(\ref{Geff-1}) that we have
\beq
\mbox{laboratory:} \;\;\; \frac{\beta^2}{K'} \leq 10^{-4} \;\; \mbox{for} \;\; \chi \sim \chi_{\rm lab} .
\label{Kp-lab}
\eeq
In this case, the screening is induced by the gravitational field of the Earth. Since
$R_K(M_{\oplus}) \simeq 2$ AU, we have at the surface of the Earth
$[R_K(M_{\oplus})/R_{\oplus}]^2 \sim 2 \times 10^9$, and Eq.(\ref{chi-Kp-1}) gives
\beq
\sqrt{-\chi_{\rm lab}} K_{\rm lab}' \sim 10^9 .
\label{chi-lab}
\eeq
This means that laboratory experiments constrain the K-mouflage model much further into the nonlinear
regime than the Cassini or Lunar Ranging probes, with typically $|\chi_{\rm lab}| \gg |\chi_{\rm s.s.}|$.
As described in Sec.~\ref{sec:Constructing-models} below, as explicit models that pass all constraints
we shall consider simple models where $K'$ converges to a large constant value $K_*$ in the nonlinear
regime $|\chi | \gg \chi_*$. Then, the constraint (\ref{Kp-lab}) on $K'$ is less stringent but close to the
Solar System one (\ref{screening-constraints-1}).}

Finally, the Lunar Laser Ranging constraint on the local rate of change of Newton's gravitational strength, Eq.~(\ref{Lunar-1}), implies
\beq
\mbox{bound on the time dependence of } \tilde{\cG} : \;\;\; \beta \leq 0.1 .
\label{lLunar-2}
\eeq

\subsection{Possible kinetic functions}
\label{sec:window}

From Eqs.~(\ref{cosmo-constraints}) and (\ref{lLunar-2}), we note that the cosmological and local (Earth-Moon system) constraints on the coupling parameter happen to
be of the same order, $\beta \lesssim 0.1$.
In terms of the kinetic function $K(\chi)$, the cosmological and small-scale constraints
apply to different branches, $\chi>0$ and $\chi<0$, respectively. Therefore, there seems to remain some freedom
in the choice of the function $K$. The main requirements are that $K'$ should be large in both limits $\chi \rightarrow \pm \infty$, which ensures screening
in both the early-time cosmology and the small-scale dynamics. However, in addition to this, we also have that in the small-scale regime, around $\chi_{\rm s.s.}<0$,
the kinetic function is very strongly constrained by the perihelion bound, Eq.~(\ref{perihelion-constraint}).

In order for the function $K(\chi)$ to satisfy the above requirements, one cannot avoid some degree of fine-tuning. A simple way to see this is to note that power-law behaviors cannot easily match the constraints. For instance, the Solar System regime, Eq.~(\ref{chi-Kp-ss}), requires a large degree of nonlinearity for $K(\chi)$, far away from the low-order expansion of Eq.~\ref{K-chi=0} which would give $K^\prime \sim 1$ and $(-\chi_{\rm s.s.}) \sim 10^{12}$ from Eq.~(\ref{chi-Kp-ss}), and would fail to satisfy
the screening criterion, Eq.~(\ref{screening-constraints-1}). This suggests
that $(-\chi_{\rm s.s.}) \gg 1$ is far in the nonlinear regime of the function $K(\chi)$,
where $K'$ is also much greater than unity. However, this is not sufficient to satisfy the perihelion constraint, Eq.~(\ref{perihelion-constraint}).
Thus, considering a power-law behavior
$K'(\chi) \sim (-\chi)^m$, with $m>0$, we have $|\chi K''| \sim K'$. The
perihelion constraint, Eq.~(\ref{perihelion-constraint}), is much stronger than the
Cassini constraint, Eq.~(\ref{screening-constraints-1}), and we obtain for $\beta \sim 0.1$
the lower bound $K' > 10^{10}$. Then, Eq.~(\ref{chi-Kp-ss}) would actually
imply $(-\chi_{\rm s.s.}) < 10^{-8}$.
Therefore, such a power-law regime would need to occur very close to the origin,
with $K'$ going from $1$ to $10^{10}$ as $\chi$ goes from $0$ to $-10^{-8}$, that is,
$\chi_* < 10^{-8}$.
This would be an extremely finely tuned kinetic function $K(\chi)$.

To achieve this quick growth of $K'(\chi)$, we may consider functions that diverge
at a finite negative value $-\chi_*<0$, such as $K'(\chi) \sim (\chi+\chi_*)^{-m}$ with
$m>0$. Then, saturating the upper bound of Eq.~(\ref{perihelion-constraint}) with
the condition of Eq.~(\ref{chi-Kp-ss}) gives $(\chi+\chi_*) \sim 10^{-10}$ (for $m=1$)
and $\chi_* \sim 10^{-8}$. Therefore, including singular kinetic functions does not
remove the need for extreme fine-tuning and again requires a very quick departure
from the low-$\chi$ expansion (\ref{K-chi=0}).

The way out is to suppress the second derivative $K''$, that is, to look for
kinetic functions such that
\beq
\chi \ll -\chi_* : \;\;\; | \chi K'' | \ll K' .
\label{small-K''}
\eeq
This means that $\ln(K')$ must grow much more slowly than $\ln(-\chi)$
for $\chi \rightarrow -\infty$. Typically, this corresponds to models where
$K'$ converges to a constant (although a logarithmic growth could also be possible),
\beq
\chi \rightarrow -\infty : \;\;\; K' \rightarrow K_*  \gg 1 ,
\label{Kp-inf}
\eeq
with the relation (\ref{chi-Kp-ss}) giving $K_* \sim 10^6/\sqrt{-\chi_{\rm s.s.}}$.
Then, the Cassini bound, Eq.~(\ref{screening-constraints-1}), implies for
$\beta \sim 0.1$:
\beq
\mbox{if  } \beta \sim 0.1 : \;\;\; K_* > 10^3  \;\;\; \mbox{and} \;\;\;
\chi_* < 10^6 ,
\label{K*-inf}
\eeq
which also ensures that the laboratory constraint (\ref{Kp-lab}) is satisfied.
On the other hand, if we require $(-\chi_{\rm s.s.}) \gtrsim 1$, so that the
transition from $K'(0)=1$ to $K_*$ does not take place in a very small interval,
to avoid extreme fine-tuning, we have the upper bound
\beq
\mbox{if  }\;\; \chi_* \gtrsim 1 : \;\;\; K_* < 10^6 .
\label{K*-sup}
\eeq
Thus, we obtain a finite range for the possible values of $K_*$. In particular,
it happens that too large values of $K'$ are ruled out if we wish to avoid too
much fine-tuning. Moreover, Eq.~(\ref{perihelion-constraint}) yields
\beq
\frac{|\chi_{\rm s.s.} K''_{\rm s.s.}|}{K'_{\rm s.s.}} \lesssim
10^{-10} K'_{\rm s.s.} < 10^{-4} ,
\label{Kss-conv}
\eeq
which means that we must have converged to the asymptotic value of
$K'$ at the $10^{-4}$ level at least.

Admittedly, there still remains some tuning, as we require that $K'(\chi)$ goes
from unity at $\chi=0$ to an asymptotic value between $10^3$ and $10^6$
at large negative $\chi$. This transition, however, does not need to be very sharp,
as it can take place over an interval that can be as large as $\chi_* \sim 10^6$.
Nevertheless, it still requires introducing a parameter $K_* \gtrsim 10^3$ for the
asymptotic constant slope of the kinetic function. We note also that this cannot be obtained using a logarithmic growth of $K'$,
which, although consistent with Eq.~(\ref{small-K''}), as $\ln(\chi_*) <  \ln(10^6)$, it is not sufficient to generate factors of order $10^3$.

\subsection{Running of the coupling $\beta$}
\label{sec:running}

The Newtonian potential of the Sun at the orbit of the Earth is
$\Psi_{\rm N}/c^2 \simeq - 10^{-8}$ (the potential of the Earth at the orbit of the
Moon gives the smaller value $-10^{-11}$). Then, Eq.(\ref{dtheta-beta}) gives the
constraint
\beq
\frac{\beta^2}{K'^2}
\left| \beta' \right|  \leq 8 \times 10^{-5} .
\eeq
For $\beta=0.1$ and $K_*=10^3$ this yields
\beq
\left| \beta' \right| < 8000 .
\label{betap-bound}
\eeq
Therefore, in contrast with the kinetic function $K(\chi)$, the bound on the perihelion
advance does not provide useful constraints on the shape of the coupling function
$A(\varphi)$ [generic functions of the form $A(\beta\varphi/M_{\rm Pl})$ would have
$\beta' \sim \beta^2 \sim 10^{-2}$].
This is due to the small prefactor $\Psi_{\rm N}/c^2$ that appears in
Eq.(\ref{dtheta-beta}).

\section{Explicit models that pass all constraints}
\label{sec:explicit}

\subsection{Constructing models}
\label{sec:Constructing-models}

\begin{figure}
\begin{center}
\epsfxsize=8.5 cm \epsfysize=5.8 cm {\epsfbox{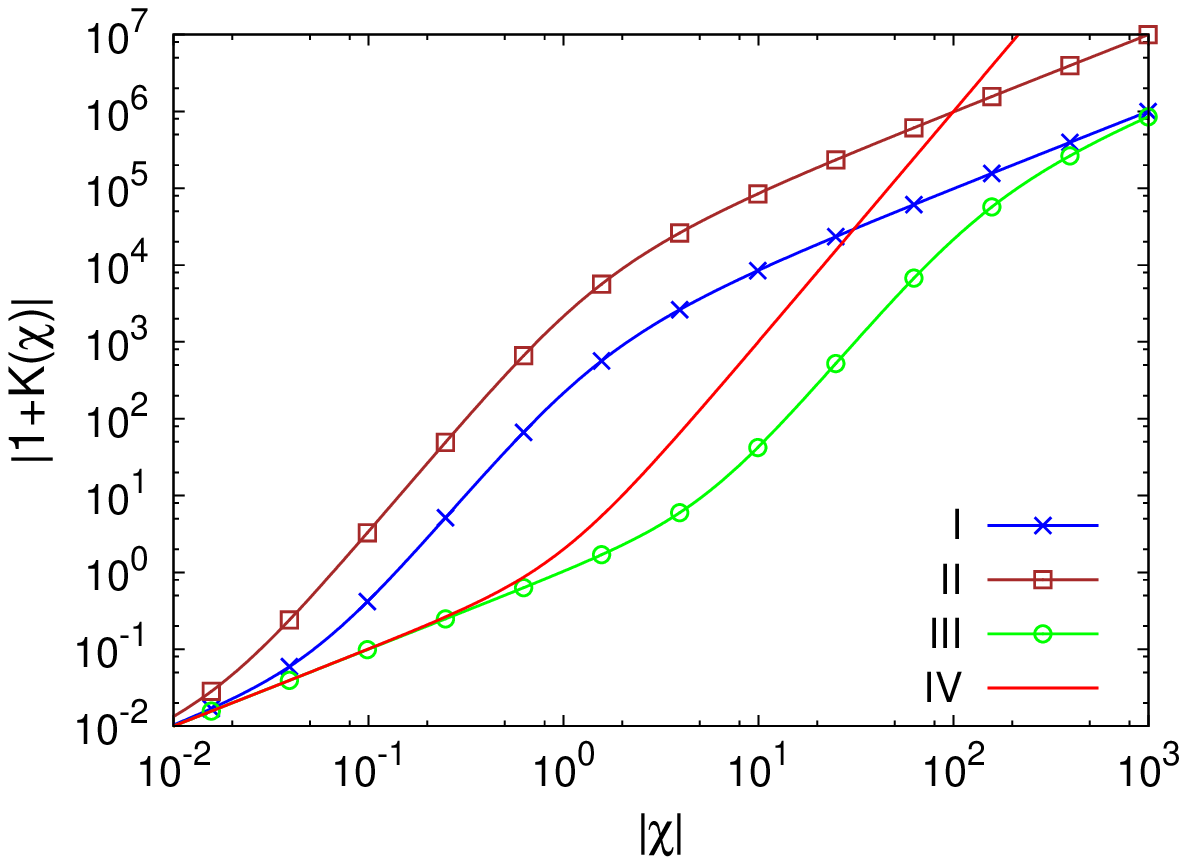}} \\
\epsfxsize=8.5 cm \epsfysize=5.8 cm {\epsfbox{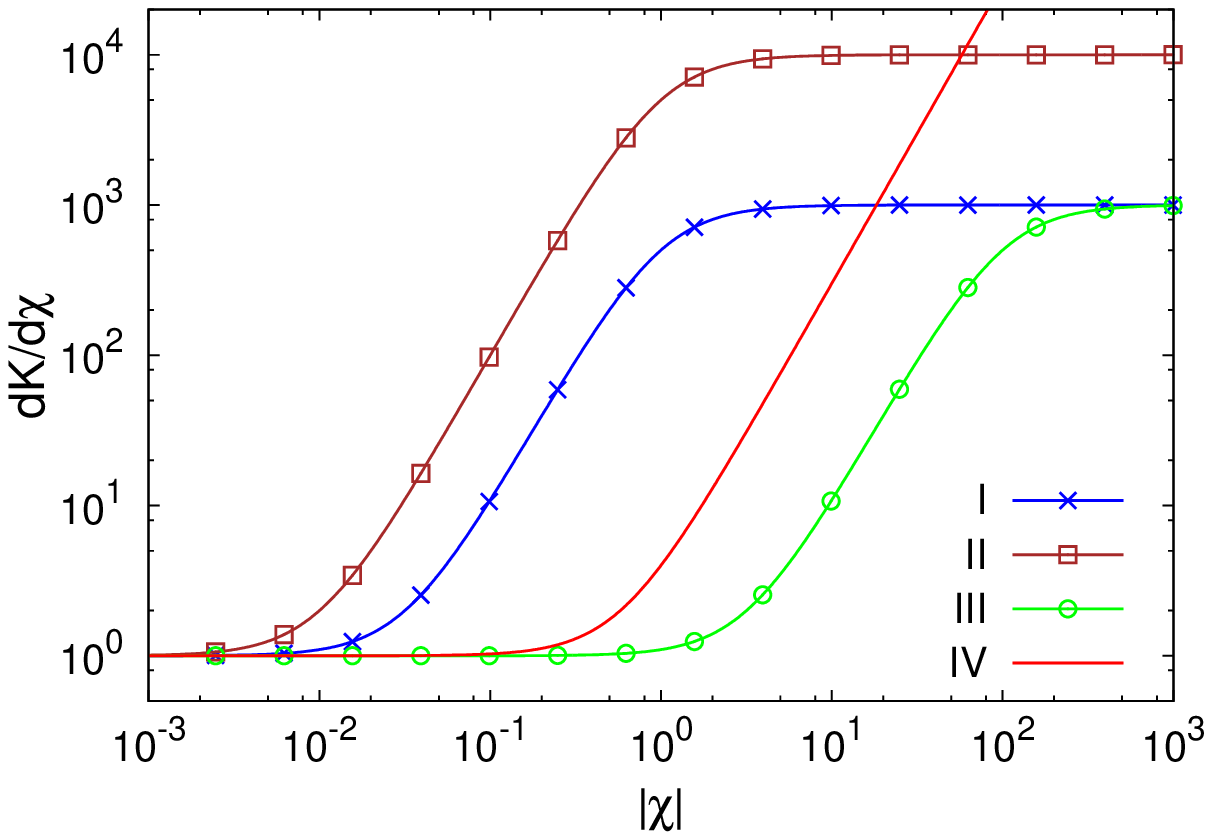}}
\end{center}
\caption{{\it Upper panel:} kinetic function $K(\chi)$ for the $\arctan$ models
of Eq.~(\ref{K-arctan-def}), with $(K_*,\chi_*)=(10^3,1)$ (I, blue and crosses),
$(K_*,\chi_*)=(10^4,1)$ (II, brown and squares), $(K_*,\chi_*)=(10^3,10^2)$ (III, green and circles),
and the cubic model of Eq.~(\ref{cubic}) with $K_0=1$ (IV, red and continuous line).
We show $1+K$, which is an odd function of $\chi$, with $1+K \geq 0$ for $\chi\geq 0$.
{\it Lower panel:} derivative $K'(\chi)$ for the same models. It is an even function of $\chi$.}
\label{fig_Kchi}
\end{figure}

A family of models which satisfy the properties obtained in Sec.~\ref{sec:window}
is given by
\be
K'= 1+K_*  \frac{\chi^n}{\chi_*^n+\chi^n} , \;\;\; n \;\; \mbox{even integer} ,
\label{def-Kp-1}
\ee
which is well defined over the full real axis (hence the choice of even integers
for the exponent $n$) and goes to $K_*$ at large $|\chi|$, with the constraints
\beq
\beta =0.1, \;\;\; K_* \geq 10^3 , \;\;\; \chi_* < \frac{10^{12}}{K_*^2} .
\label{K*def}
\eeq
This ensures that the Cassini bound (\ref{screening-constraints-1}) is satisfied
with $|\chi_{\rm s.s.}| >\chi_*$ from Eq.~(\ref{chi-Kp-ss}), {as well as
the laboratory constraint (\ref{Kp-lab}).}
Then, the perihelion constraint (\ref{perihelion-constraint}) yields
\beq
\chi_* < \left( \frac{K_*}{n} 10^{-10} \right)^{1/n} \frac{10^{12}}{K_*^2} .
\label{perihelion-2}
\eeq
As seen in Eq.~(\ref{K*-sup}), we should have $K_* < 10^6$ if we wish to avoid
tuning $\chi_*$ to values that are smaller than unity. This agrees with
Eq.~(\ref{perihelion-2}), which in such cases is more stringent than
the last equation in (\ref{K*def}).
In particular, this gives:
\beq
n =2 : \;\; \chi_* < \frac{10^7}{K_*^{3/2}}  , \;\;\;
n \rightarrow \infty: \;\; \chi_* < \frac{10^{12}}{K_*^2} .
\label{n=2-perihelion}
\eeq

In the remainder of this section we focus on the simplest model with $n=2$, which corresponds to
\beq
K(\chi) = -1 + \chi +K_* [ \chi - \chi_* \arctan(\chi / \chi_*) ] ,
\label{K-arctan-def}
\eeq
with the low-$\chi$ expansion
\beq
\chi \rightarrow 0 : \;\;\; K(\chi) = -1 + \chi + \frac{K_*}{3} \frac{\chi^3}{\chi_*^2}
- \frac{K_*}{5} \frac{\chi^5}{\chi_*^4} + ...
\label{low-chi-1}
\eeq
We consider the following three models,
\beqa
{\rm (I)} : && K_*= 10^3 , \;\;\; \chi_*= 1 , \label{I-def} \\
{\rm (II)} : && K_*= 10^4 , \;\;\; \chi_*= 1 , \\
{\rm (III)} : && K_*= 10^3 , \;\;\; \chi_*= 10^2, \label{III-def} 
\eeqa
all with $\beta=0.1$ and which satisfy the small-scale constraints of Eqs.~(\ref{K*def}) and (\ref{n=2-perihelion}),
as well as the cosmological constraints.
In particular, in each case we have
\beqa
{\rm (I)}\!\! & : & \chi_{\rm s.s.} \sim -10^6 , \;\; \beta^2_{\rm per} <  500 , \;\;
\frac{c_s^2}{c^2} - 1 \sim 10^{-12} , \;\;\;\;\;\; \label{cs-I} \\
{\rm (II)} \!\! & : & \chi_{\rm s.s.} \sim -10^4 , \;\; \beta^2_{\rm per} <  0.5 , \;\;
\frac{c_s^2}{c^2} - 1 \sim 10^{-8} , \\
{\rm (III)} \!\! & : & \chi_{\rm s.s.} \sim -10^6 , \;\; \beta^2_{\rm per} <  0.05 , \;\;
\frac{c_s^2}{c^2} - 1 \sim 10^{-8} , \label{cs-III}
\eeqa
where $\chi_{\rm s.s.}$ is obtained from Eq.(\ref{chi-Kp-ss}) and $\beta_{\rm per}$
is the upper bound from Eq.(\ref{perihelion-constraint}). Thus, $\beta=0.1$
is consistent with the perihelion constraint for these three models.
We can also note that the speed of scalar waves given by Eq.(\ref{cs-def})
is always very close to the speed of light (in the Solar System).
Therefore, superluminality in the scalar sector is highly suppressed. 

We show these kinetic functions in Fig.~\ref{fig_Kchi}, as a function of $|\chi|$.
The models (I) and (II) correspond to a late transition from the $K' \simeq K_*$ to
$K' \simeq 1$ regimes, with two possible values for $K_*$.
Therefore, during most of the cosmological evolution $K' \gg 1$,
which means that the scalar field is screened and departures from $\Lambda$-CDM are small.
The model (III) corresponds to an early transition, so that in a large range of redshifts
we have $K' \simeq 1$ and deviations from $\Lambda$-CDM are of order $\beta^2=10^{-2}$.
These behaviors are explicitly shown in the middle panel of Fig.~\ref{fig_chi_z}.

For simplicity we consider the simple rational function of Eq.~(\ref{def-Kp-1}), which is
even and leads to the same behavior in the two regimes $\chi \rightarrow \pm \infty$.
However, there is a great freedom in the positive range $\chi > 0$, the only
constraint being that $K'(\chi) \gg 1$ for $\chi \gg 1$.
For instance, we could add to Eq.~(\ref{K-arctan-def})
any function $K_+(\chi)$ that is negligible for $(-\chi) \gg 1$ and satisfies
$K_+'(0)=0$, $K_+'(\chi)+K_* \gg 1$ at large $\chi$, such as $K_+(\chi) = \exp(\chi^3)$.

For comparison with previous works \cite{Brax:2014a,Brax:2014b,Brax:2014c,Barreira:2014a},
we also consider the purely cubic model
\beq
{\rm (IV)} : \;\;\; K(\chi)=-1 +\chi + K_0 \chi^3 .
\label{cubic}
\eeq
This can also be seen as an effective model for moderate values of $\chi$,
probed by cosmology, while leaving the large negative regime $(-\chi) \gg 1$
unspecified, where the function $K(\chi)$ needs to be modified as described above
to satisfy Solar System constraints.
This model is also shown in Fig.~\ref{fig_Kchi}, for the case $K_0=1$.
For most of the cosmologically relevant values of $\chi$, $0\leq \chi \lesssim 10$, the phenomenology of model (IV) lies between that of models (I) and (III).

\subsection{Theoretical consistency}
\label{sec:consistency}

We have seen above that the models (I)-(III) satisfy the quantitative constraints
associated with cosmological and Solar System tests.
In addition, they satisfy the generic theoretical consistency requirements.
The four models (I)-(IV) have an even derivative $K'(\chi)>0$ and the
functions $W_{\pm}(y)=y K'(\pm y^2/2)$ are monotonically increasing up to infinity
over $y\geq 0$.
As shown in Refs.~\cite{Brax:2014a,Brax:2014c}, this ensures that these models 
are well behaved. More specifically, a static scalar field profile exists for any matter 
density profile (branch $W_-(y)$, see Eq.~(\ref{chi-Kp})) and the background 
Klein-Gordon equation can be integrated up to arbitrarily high redshifts where
$\bar\rho\rightarrow\infty$.
Moreover, there are also no ghosts nor small-scale instabilities.
This corresponds to the constraints 
(\ref{cosmo-constraints-Kp-Ks})-(\ref{cosmo-constraints-Wp}) and
(\ref{static-constraints-Kp-Ks})-(\ref{static-constraints}).
When the function $K'(\chi)$ is even these two sets of constraints actually coincide.

As noticed in Eq.(\ref{cs-def}), the speed of scalar waves around static backgrounds
is greater than the speed of light. As seen from Eqs.(\ref{cs-I})-(\ref{cs-III}),
it is actually extremely close to $c$ in the highly nonlinear regime, which applies
to the Solar System.
Superluminality is sometimes associated with possible violations of causality,
although such pathological behaviors can also be encountered within General
Relativity (for exact solutions such as the G$\rm{\ddot{o}}$del metric that do not describe
realistic metrics), see the discussions in \cite{Babichev:2007dw} and 
\cite{Wald1984}.
We argue in the appendix that space-time does not have closed time-like loops 
for realistic K-mouflage models, in particular for the models (I)-(III) of 
Eqs.(\ref{I-def})-(\ref{III-def}).
Therefore, the small superluminality in the Solar System does not lead to
causality problems.

\subsection{Cosmological evolution}
\label{sec:Cosmological-evolution}

\begin{figure}
\begin{center}
\epsfxsize=8.5 cm \epsfysize=5.8 cm {\epsfbox{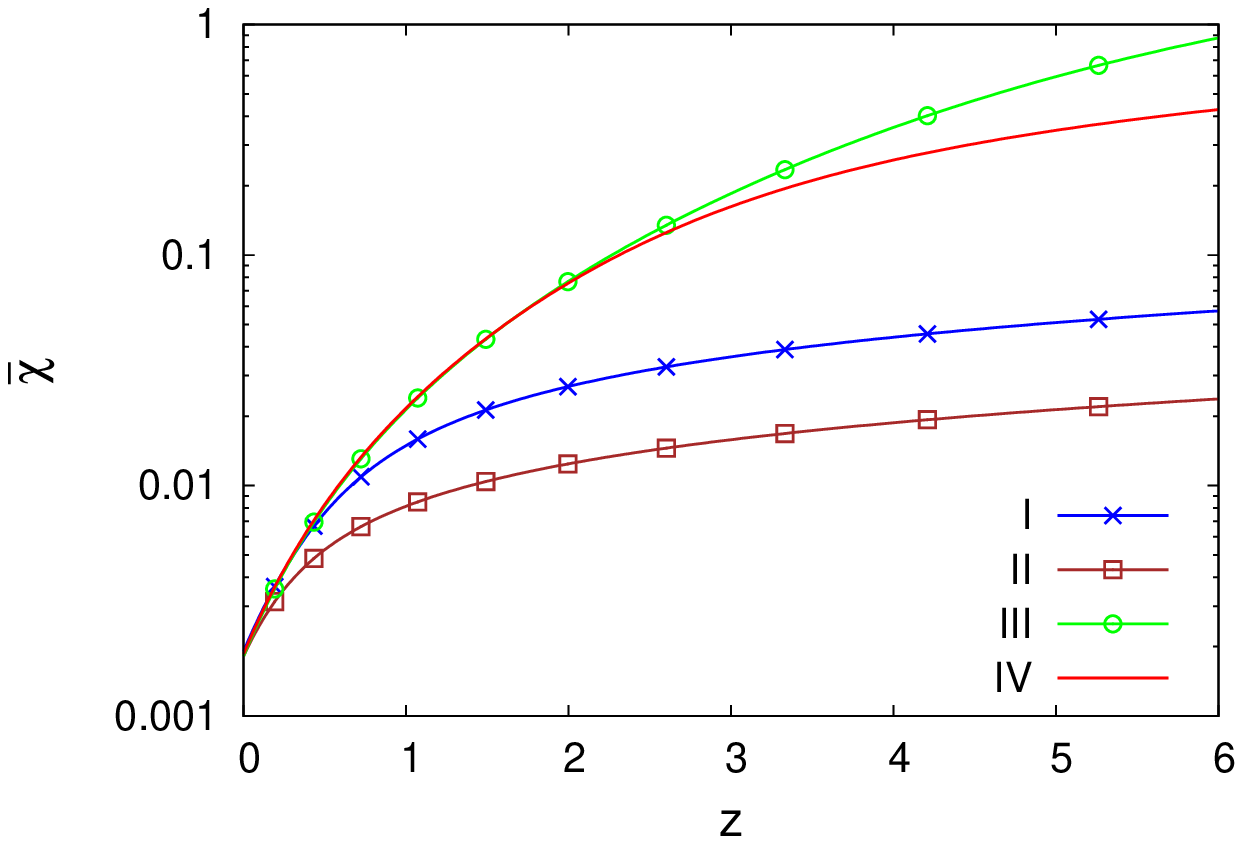}} \\
\epsfxsize=8.5 cm \epsfysize=5.8 cm {\epsfbox{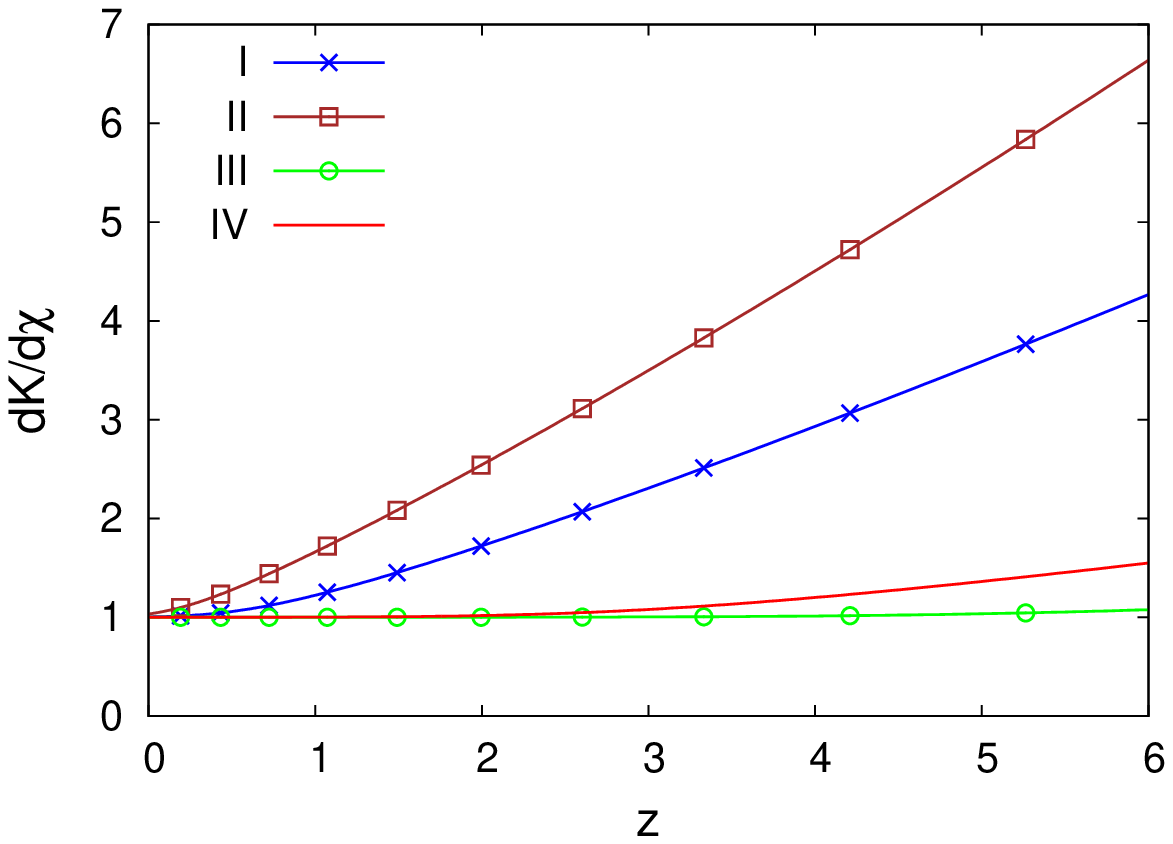}} \\
\epsfxsize=8.5 cm \epsfysize=5.8 cm {\epsfbox{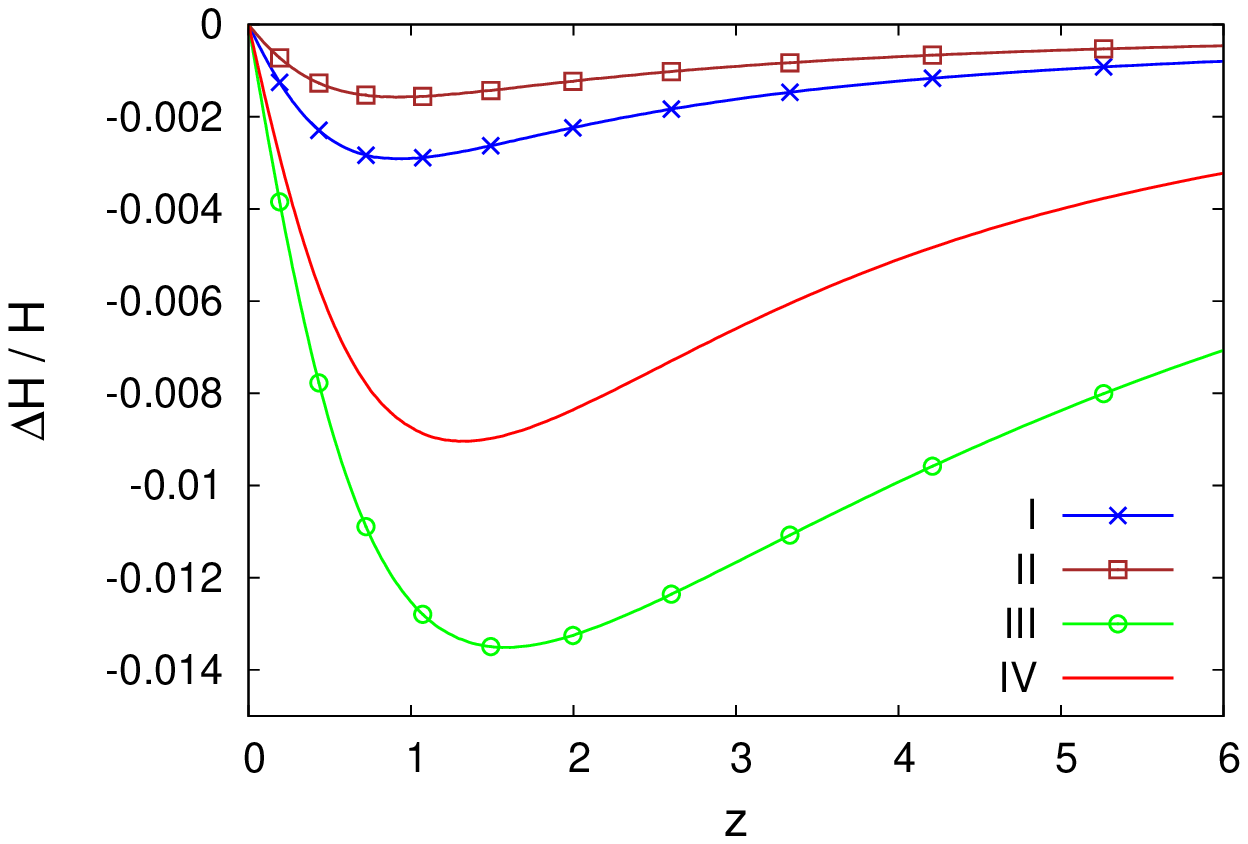}}
\end{center}
\caption{{\it Upper panel:} variation of $\bar\chi(z)$ as a function of the redshift for the models
of Fig.~\ref{fig_Kchi}. All models have $\beta=0.1$
Notice the universal behavior of $\bar\chi$ at small redshift given by (\ref{dot-phi-chi}) with $K'\simeq 1$.
{\it Middle panel:} variation of $\bar{K}'=\dd K/\dd\chi(\bar\chi)$ as a function of the redshift for the same models.
{\it Lower panel:} relative deviation of the Hubble rate from the $\Lambda$-CDM prediction for the same
models.}
\label{fig_chi_z}
\end{figure}

\begin{figure}
\begin{center}
\epsfxsize=8.5 cm \epsfysize=5.8 cm {\epsfbox{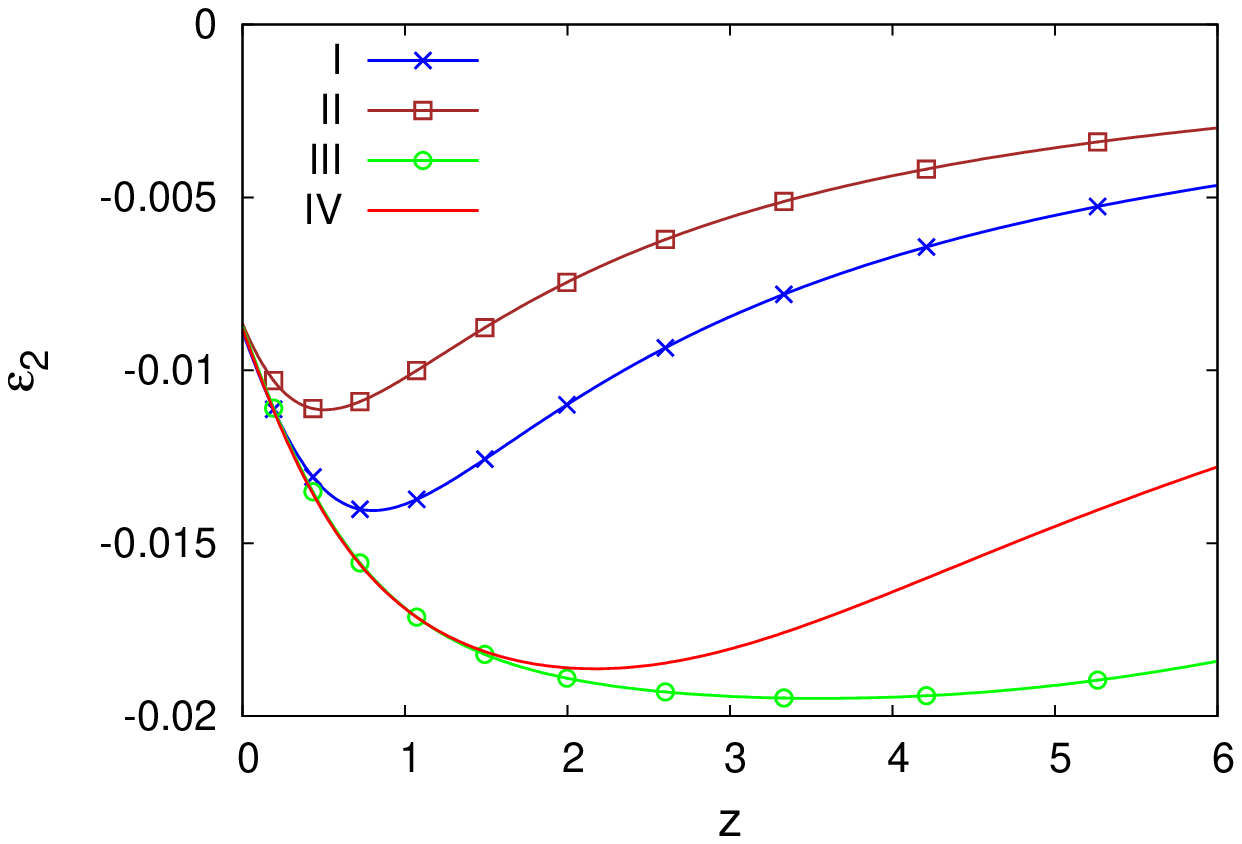}} \\
\epsfxsize=8.5 cm \epsfysize=5.8 cm {\epsfbox{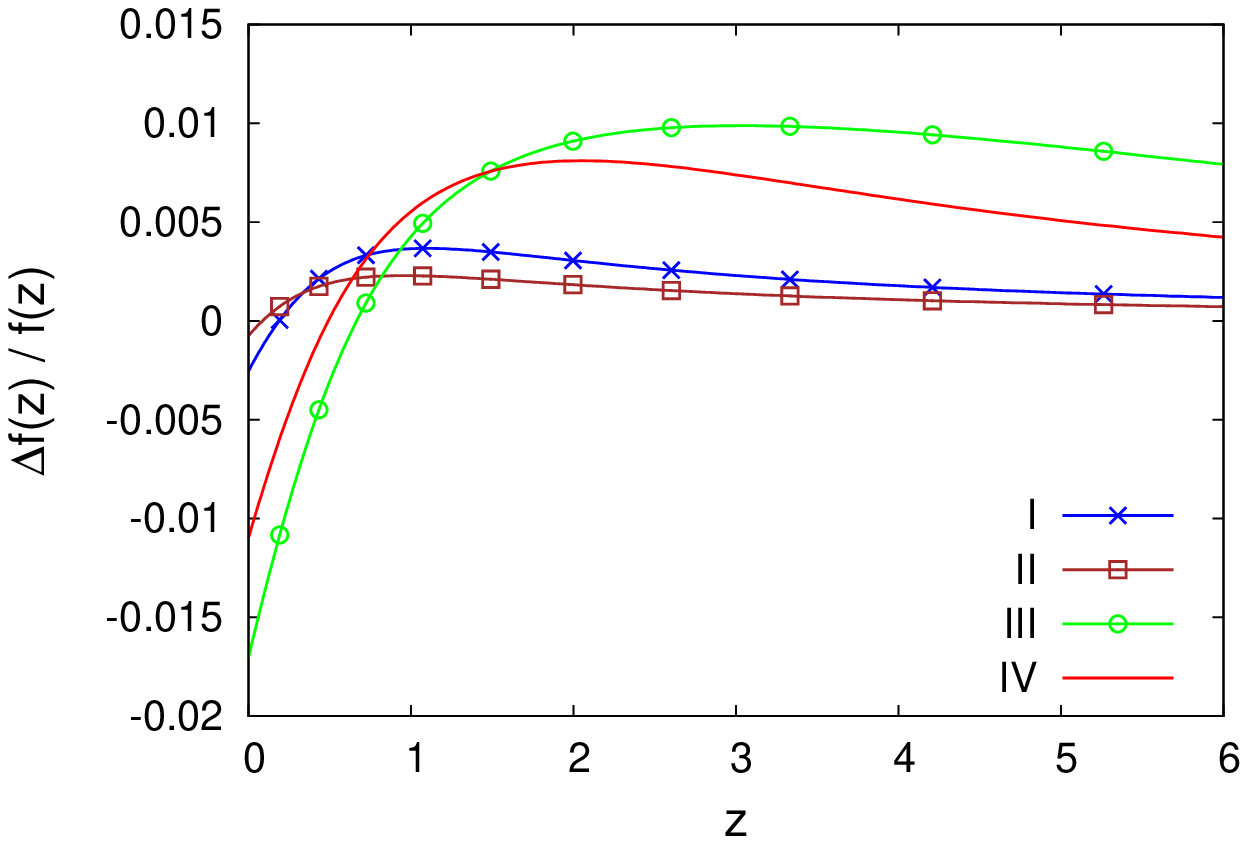}}
\end{center}
\caption{{\it Upper panel:} variation of $\epsilon_2(z)$ as a function of the redshift for the models of
Fig.~\ref{fig_Kchi}.
{\it Lower panel:} relative deviation of the growth factor from the $\Lambda$-CDM prediction for the same
models.}
\label{fig_eps_z}
\end{figure}

\begin{figure}
\begin{center}
\epsfxsize=8.5 cm \epsfysize=5.8 cm {\epsfbox{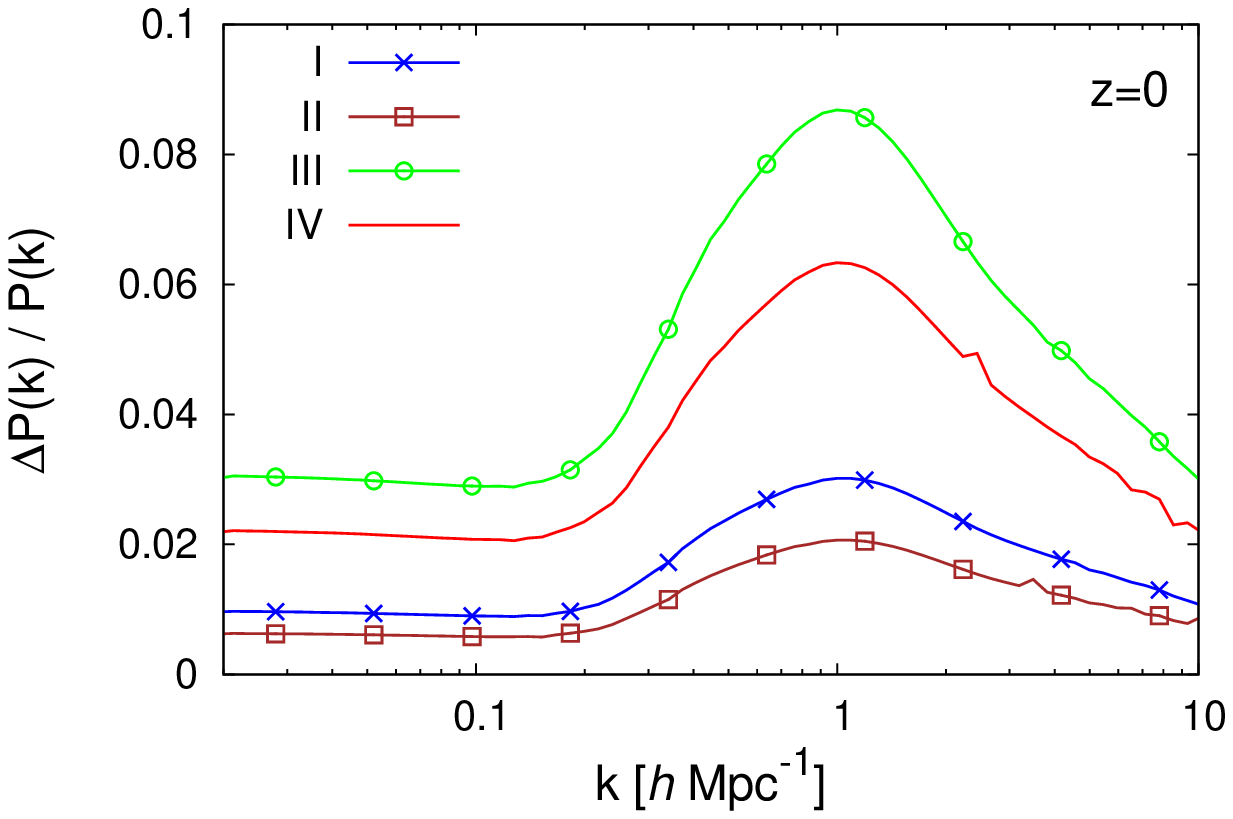}} \\
\epsfxsize=8.5 cm \epsfysize=5.8 cm {\epsfbox{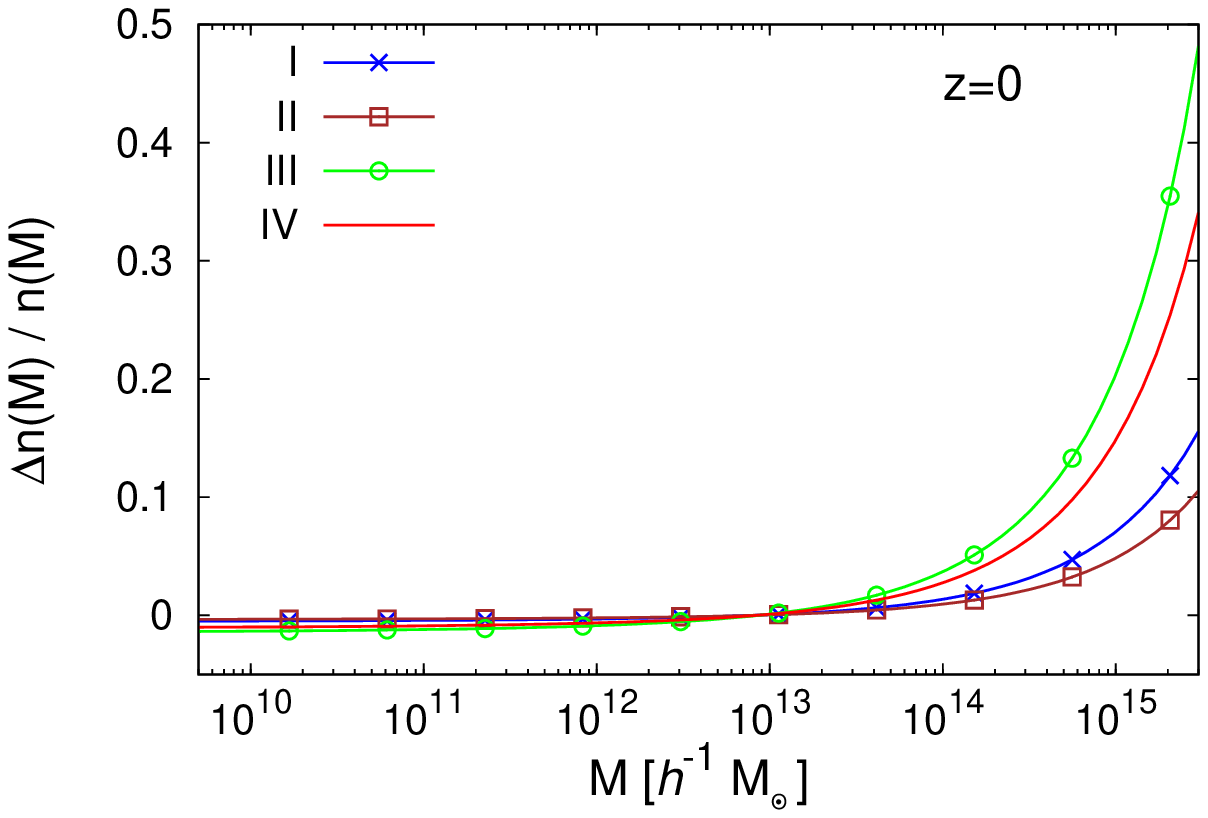}}
\end{center}
\caption{{\it Upper panel:} relative deviation of the power spectrum from $\Lambda$-CDM for the
models of Fig.~\ref{fig_Kchi}, at $z=0$.
{\it Lower panel:} relative deviation of the halo mass function for the same models, at $z=0$.}
\label{fig_Pk}
\end{figure}

From the Klein-Gordon equation (\ref{KG-1}), one can show \cite{Brax:2014a}
that the time derivative of the cosmological background value $\bar\varphi$ of the scalar field
scales with time as
\beq
\dot{\bar\varphi} \sim - \frac{\beta}{M_{\rm Pl} K'} \, \rho t , \;\;\; \mbox{and} \;\;\;
\bar\chi \sim \frac{\beta^2}{K'^2} \frac{H^2}{H_0^2} .
\label{dot-phi-chi}
\eeq
At late times we have
\beq
z \rightarrow 0 : \;\; \bar{K}' \simeq 1, \;\; \bar\chi \sim \beta^2 = 10^{-2} , \;\;\;
\bar\rho_{\varphi} \simeq \frac{\Omega_{\Lambda 0}}{\Omega_{\rm m 0}} \bar\rho ,
\eeq
where $\Omega_{\Lambda 0}$ and $\Omega_{\rm m 0}$ are the dark energy and
matter cosmological parameters today. At early times we have, for
the models (I), (II), and (III),
\beq
z \gg 1 : \;\;\; \bar{K}' \simeq K_* , \;\; \bar\chi \sim \frac{\beta^2}{K_*^2} \frac{H^2}{H_0^2} ,
\;\;\; \bar\rho_{\varphi} \sim \frac{\beta^2}{K_*} \bar\rho .
\eeq
Thus, in the early matter era, the ratio $\bar\rho_{\varphi}/\bar\rho$ goes to a small
constant, which in our case is constrained by the Cassini bound
Eq.~(\ref{screening-constraints-1}) to be smaller than $10^{-5}$
(this is because we assume the two regimes $\chi\rightarrow \pm\infty$ have the same
functional form).
Thus, there remains a very small residual fraction of the scalar field energy density.
For the model (IV) of Eq.~(\ref{cubic}), where $K'(\chi)$ goes to infinity for
$\chi\rightarrow+\infty$, $\bar\rho_{\varphi}/\bar\rho \rightarrow 0$ at high redshift.
In all cases, $\bar\rho_{\varphi}/\bar\rho_{\rm rad}$ goes to zero at high redshift in the
radiation-dominated era.

At low redshifts, where $\bar\chi$ falls below $\chi_{\rm can.} = \chi_*/\sqrt{K_*}$,
we have $\bar{K}' \simeq 1$ and the model behaves like a canonically normalized scalar field
coupled to matter in the presence of a cosmological constant.
At higher redshifts, the nonlinear structure of the scalar field Lagrangian comes into play
and $\bar{K}'$ increases until it reaches $K_*$ when $\bar\chi$ is of order
$\chi_*$. Then, the scalar field shows a K-mouflage-type screening, due to the large
derivative prefactor $K'$ that freezes the fluctuations of the scalar field.
In terms of uniform background values, this screening still leads to the suppression
of $\bar\chi$ and $\bar\rho_{\varphi}$ by factors $1/\bar{K}'^2$ and $1/\bar{K}'$.

The deviations from General Relativity and $\Lambda$-CDM predictions are mostly sensitive
to the ratio $\beta^2/K'$. Therefore, they should increase from models (II) to (I) to (III).
Indeed, in models (I) and (II) the derivative $K'$ quickly reaches the large value $K_*$, because
$\chi_*=1$, while in model (III) it remains of order unity over a large interval around $\chi=0$,
because $\chi_* \gg 1$. The deviations from $\Lambda$-CDM are larger for model (I) than model (II) because of the
smaller value of $K_*$ in the former.

In Fig.~\ref{fig_chi_z}, we plot the redshift evolution of $\bar\chi$ and $\bar{K}'$, and the deviations
of the Hubble rate from $\Lambda$-CDM. The figure shows that, as expected, the deviations from $\Lambda$-CDM increase from models (II) to (I) to (III).
Also, in agreement with Fig.~\ref{fig_Kchi}, the predictions of the cubic model (IV) lie in between the arctan models (I) and (III), as it gives an intermediate
growth of $\bar{K}'$ as $\bar\chi$ increases at higher $z$. At $z\rightarrow 0$ all models give the same value
for $\bar\chi$, in agreement with Eq.~(\ref{dot-phi-chi}), because $\bar\chi \sim \beta^2 \ll 1$ and
$\bar{K}' \simeq K'(0)=1$.
The deviations from $\Lambda$-CDM for the Hubble rate peak around $z \sim 1$. All models are normalized to the same Hubble constant $H_0$ today and they also recover the
same Einstein-de Sitter expansion at high redshift in the matter era, before dark energy becomes
important.

The upper panel of Fig.~\ref{fig_eps_z} shows that the four models satisfy the Lunar Ranging constraint on
$\dd \ln \tilde{\cG}/\dd \tilde{t}$ as the values of $\epsilon_2$ are below $0.01$, in agreement with Eq.~(\ref{eps2-1}).
This, however, does not prevent the linear growth rate $f= \dd\ln D_+/\dd\ln a$ today 
from being different in the different models, as shown in the lower panel. The deviations from $\Lambda$-CDM of the linear fluctuations follow the same pattern as for the background cosmology.  We also find that a maximal deviation of order $2\beta^2$ is obtained for model (III) with a large value of $\chi_*$, which behaves like a free scalar field coupled to matter in the recent past. The greatest deviations from $\Lambda$-CDM, at fixed $\beta$, would be obtained for a model where $K' \simeq 1$ over all relevant redshifts. The results from such a model would be similar to those of model (III).

In Fig.~\ref{fig_Pk}, we show the relative deviations from $\Lambda$-CDM of the nonlinear matter power spectrum and the halo mass function. The method used for the computation of the power
spectrum was presented in \cite{Brax:2014b}. It combines one-loop perturbation theory
and a halo model and it has been tested against numerical simulations \cite{BraxPV2013} of other
modified gravity scenarios such as $f(R)$, dilaton and symmetron models.
The power spectrum deviates on large scales by a constant which can reach a few percent. This reflects the boost in the linear growth of structure depicted in the lower panel of Fig.~\ref{fig_eps_z}. In the nonlinear regime, mode coupling helps to boost the deviations from standard $\Lambda$-CDM even further. In the figure, the relative difference starts to decrease after $k \sim 1 \; h/{\rm Mpc}$, which is a consequence of having used the same halo concentration parameters in the halo model formalism \cite{Brax:2014b}. In reality, the K-mouflage field should also affect the concentration of dark matter haloes, which should translate in modifications of the small-scale clustering power as well. The enhanced gravitational strength should lead to a deepening of the gravitational potentials of the haloes, making them therefore more concentrated (see e.g.~\cite{Barreira:2014zza, Barreira:2014kra} for an example of this in the context of Galileon and Nonlocal models of gravity, respectively). This would increase the relative difference of $P_k$ at high-$k$ values. On the other hand, the enhanced forces may also make particles inside haloes  move faster. This can cause the halo to expand slightly, and therefore, to become less concentrated (see e.g.~\cite{2011PhRvD..83b4007L} for an example of this in coupled quintessence models, which are similar to unscreened K-mouflage models). A detailed investigation of this effect involves running dedicated $N$-body simulations which is left for future work.

For the halo mass function, we take into account the effect of K-mouflage
through the spherical collapse model \cite{Brax:2014b}.
The amplification of gravity by the fifth force implies that a smaller initial density
fluctuation $\delta_{L\rm i}$ at a high redshift $z_{\rm i}$ is needed to produce
a collapsed halo of a given mass $M$ at low $z$, as compared with the
$\Lambda$-CDM reference. Moreover, such objects are less rare because of the
faster growth of structures (e.g., of the matter power spectrum).
Both effects amplify the large-mass tail of the halo mass function, as seen
in the lower panel of Fig.~\ref{fig_Pk} [the mass function is written in terms
of the usual scaling variable $\nu=\delta_L/\sigma(M)$, with the characteristic
Gaussian tail $n(M) \sim e^{-\nu^2/2}$ at large mass, and the fifth force implies
both a smaller critical threshold $\delta_L$ and a greater rms density fluctuation
$\sigma(M)$]. The mass function is slightly decreased at low masses because of
mass conservation, as the integral over the halo mass function is normalized so as to
give back the mean density of the Universe.
The lower panel of Fig.~\ref{fig_Pk} shows that K-mouflage models can exhibit a $5-20\%$ boost in the number density of cluster mass haloes, $10^{14} M_{\odot}/h \lesssim M \lesssim 10^{15} M_{\odot}/h$. This may offer a clear enough signal to be observable with future surveys.

\section{conclusion}

We have determined the conditions for K-mouflage models of gravity to satisfy the stringent Solar System tests of gravity, whilst remaining sufficiently different from standard $\Lambda$-CDM, and hence, cosmologically interesting. In particular, we have used the results from Cassini on the amplitude of fifth forces in the Solar System, and bounds from Lunar Laser Ranging experiments on the anomalous perihelion and the rate of change of the gravitational strength in the Earth-Moon system.

We showed that the conformal coupling strength of the K-mouflage field to matter is constrained as $\beta \lesssim 0.1$ (cf.~Eq.~(\ref{Lunar-1})), by the condition that the time variation of Newton's gravitational strength in the Jordan frame is compatible with the Lunar Ranging bounds. This constraint is independent of the detailed functional form of the K-mouflage Lagrangian density term $K(\chi)$ [cf.~Eq.~(\ref{K-def})], and is also tighter than the bounds coming from BBN (cf.~Eq.~(\ref{BBN-2})). By focusing on static configurations, which correspond to the branch $\chi < 0$ of the function $K(\chi)$, we have seen that the perihelion bound is more stringent than the Cassini result in constraining the functional form of $K(\chi)$. In particular, for K-mouflage models to remain compatible with these tests, any nonlinear terms in $K(\chi < 0)$ should be highly suppressed (cf.~Eq.~(\ref{perihelion-constraint})). For instance, models for which $K(\chi) \propto \chi^3$ will fail to meet the Solar System requirements.
On the other hand, the shape of the coupling function $A(\varphi)$ is not strongly
constrained.

We have presented several explicit models that satisfy these Solar System and
cosmological bounds as well as generic theoretical consistency conditions.
In particular, these models have well-defined solutions up to arbitrarily high redshift,
in the cosmological context, and for any matter density profile, in the small-scale
context. There are no ghosts nor small-scale instabilities. Even though scalar
waves can propagate at a speed that is slightly greater than the speed of light
around small-scale static backgrounds (with a relative difference of only $10^{-8}$ or less
in the Solar System), there are no closed time-like loops nor causality problems, 
from the Solar System to cosmological scales.

One of our main results is that, despite the constraints on $\beta$ and $K(\chi)$ from the Solar System tests, one is still able to find a family of $K(\chi)$ functions (cf.Eq.~(\ref{def-Kp-1})) that has interesting and potentially testable cosmological predictions. We have investigated the main cosmological aspects of the models characterized by Eq.~(\ref{def-Kp-1}). For a set of illustrative cosmological parameters, we have seen that the linear growth of large-scale structures can be boosted by a few percent by the present day (cf.~Figs.~\ref{fig_eps_z} and \ref{fig_Pk}). Our results from semi-analytical models of structure formation also show that this difference gets amplified on smaller scales, where the evolution of the matter density field becomes nonlinear. Moreover, the expected number density of cluster mass haloes shows also a $5-20\%$ enhancement, relative to $\Lambda$-CDM (cf.~Fig.~\ref{fig_Pk}). Another interesting aspect of these models is that their expansion history
can be slightly different from the $\Lambda$-CDM scenario, with deviations at the percent
level or slightly below that may be constrained by observations (cf.~Fig.~\ref{fig_chi_z}).
This is different from the cases of DGP and/or $f(R)$ models of gravity, where the expansion can follow the $\Lambda$-CDM scenario
up to very high accuracy or even exactly.
This means that the parameter space of these models can be constrained by the position of the acoustic peaks of the CMB temperature power spectrum, as investigated already in \cite{Barreira:2014a}, but for models that fail Solar System tests.

{To summarize, the models we have built up are predictive and distinguishable from other alternatives to $\Lambda$-CDM, in the perspective of future experiments such as Euclid \cite{2011arXiv1110.3193L} and LSST \cite{2012arXiv1211.0310L}.   In the future, 21-cm intensity mapping both during ~\cite{Brax:2012cr} and after the completion of the reionization ~\cite{Hall:2012wd,Santos:2015gra,Zhao:2015wqa} will open new windows to test modified gravity and will help in discriminating between models and in constraining further the shape of the K-mouflage function $K(\chi)$.  Future 21-cm surveys such as with the Square Kilometre Array will probe the Universe's expansion up to higher redshifts and the matter power spectrum down to smaller scales, especially in the range $2 \lesssim z \lesssim 8$ of interest for the K-mouflage model.  }

In future work, we believe it would be of interest to perform more focused studies of the cosmological constraints in these models, by following, for instance, the line of work of \cite{Barreira:2014a}. It would also be interesting to study more accurately the predictions for nonlinear structure formation by running $N$-body simulations. Such studies should provide a clearer picture of how these types of modifications to gravity can impact on several cosmological observables, which should help in the interpretation of the results from future observational missions.

\begin{acknowledgments}

A. B. acknowledges support by FCT-Portugal through grant SFRH/BD/75791/2011.
Ph. B. acknowledges partial support from the European Union FP7 ITN INVISIBLES (Marie Curie
Actions, PITN- GA-2011- 289442) and from the Agence Nationale de la Recherche under contract
ANR 2010 BLANC 0413 01.
The work of S. C. is supported by the \textit{mandat de retour} program of the Belgian Science Policy (BELSPO).
B. L. is supported in part by STFC consolidated grant Np.~ST/L00075X/1.
P. V. acknowledges support from the French Agence Nationale de la Recherche
under Grant ANR-12-BS05-0002.

\end{acknowledgments}

\appendix

\section{Superluminality and Causality}

In this appendix, we elaborate on the superluminality of scalar perturbations and its link with causality. We have seen in the main text that K-mouflage models passing the Solar System
tests are such that scalar perturbations around a static background propagate with a speed greater than the speed of light. This may cause instabilities and in particular a loss
of causality with signals being transmitted to the past, i.e. the existence of time-like closed curves. This issue was tackled for K-essence models in \cite{Babichev:2007dw}. We follow a similar method here.
Let us first expand the K-mouflage action to second order in $\pi$, where 
$\varphi=\bar\varphi + \pi$. Here $\bar\varphi(\vx,t)$ is a background configuration
that may depend on scale and time [in the small-scale static regime $\bar\varphi(\vx)$
only depends on position, whereas in the large-scale cosmological regime
$\bar\varphi(t)$ is the homogeneous cosmological background].
We also denote 
$\bar\chi=-\bar{g}^{\mu\nu}\pl_{\mu}\bar\varphi \pl_{\nu}\bar\varphi/2\cM^4$,
$\bar K'= K'(\bar \chi)$ and $\bar K''= K''(\bar\chi)$. 
The second order part of the action reads
\be
S_{2}= \frac{1}{2}\int d^4x \sqrt{-g}  \left[ -\bar K' \pl^{\mu}\pi \pl_{\mu}\pi 
+ \frac{\bar K''}{{\cal M}^4} (\pl^{\mu}\bar\varphi \pl_{\mu} \bar\pi)^2 \right] .
\ee
It is convenient to define the disformal metric
\beq
G^{\mu\nu} = \gamma^{-1} \left( \bar K' g^{\mu\nu} 
- \frac{\bar K''}{\cM^4} \pl^{\mu}\bar\varphi \pl^{\nu}\bar\varphi \right)
\eeq
with
\beq
\gamma = (\bar K')^{3/2} (\bar K'+2\bar\chi \bar K'')^{1/2} > 0 .
\label{gamma-def}
\eeq
Note that thanks to the properties (\ref{cosmo-constraints-Kp-Ks}) and
(\ref{static-constraints-Kp-Ks}), the K-mouflage scenarios that we consider
have $K'>0$ and $K'+2\chi K''>0$ over all $\chi$, so that the metric
$G^{\mu\nu}$ is well defined.
Defining the inverse matrix $G_{\mu\nu}$ by 
$G_{\mu\alpha}G^{\alpha\nu}=\delta_{\mu}^{\nu}$ (i.e., $G_{\mu\nu}$ is not given by
$g_{\mu\alpha}g_{\nu\beta}G^{\alpha\beta}$) and the determinant
$G=\det(G_{\mu\nu})$, we have
\beq
G= \gamma^2 \, g ,
\label{det-G-munu}
\eeq
where $g=\det(g_{\mu\nu})$, and the second-order action can also be written as
\be
S_{2}= \frac{-1}{2} \int d^4 x \, \sqrt{-G} \, G^{\mu\nu} \, \partial_\mu \pi \partial_\nu \pi .
\label{S2-Gdisf}
\ee
The disformal metric $G_{\mu\nu}$ is the metric felt by the scalar perturbation
and we can check from Eq.(\ref{det-G-munu}) that it is Lorentzian.
Therefore, initial-value problems for $\pi$ are well posed on any smooth
space-like Cauchy surface $\Sigma$ for the metric $G_{\mu\nu}$, and the solution
is unique and propagates causally (see Sec.10 in \cite{Wald1984}).
In the static case, we retrieve that the Klein-Gordon equation is hyperbolic with
a propagation speed $c_s$ given by Eq.(\ref{cs-def}) (for small wavelengths),
whereas around the cosmological background we recover the
propagation speed $\bar{c}_s$ given by Eq.(\ref{bar-cs}) (for high frequencies).

In general, the propagation of $\pi$ occurs in the disformal metric $G_{\mu\nu}$. 
Space-time equipped with the metric $G_{\mu\nu}$ is stably causal, i.e. there are no 
time-like closed loops (including for infinitesimal deviations from the metric 
$G_{\mu\nu}$), provided there exists a globally defined function $f$ on all space-time 
which is time-like, i.e. $G^{\mu\nu} \partial_\mu f \partial_\nu f <0$ \cite{Wald1984}.
Following \cite{Babichev:2007dw}, we look for a ``global time'' $f$ that applies
to both geometries $g_{\mu\nu}$ and $G_{\mu\nu}$ and thus guarantees the
absence of closed causal loops. A simple choice is to choose the cosmic time
$t$, which clearly satisfies the required property for the metric $g_{\mu\nu}$.
Considering the Newtonian gauge, which describes all systems that we
study in this paper, from the cosmological background and perturbative regime 
down to the Solar System,
\beq
\dd s^2 = g_{\mu\nu} \dd x^\mu \dd x^\nu = - (1+2\Psi_{\rm N}) \dd t^2 
+ a^2(t) (1-2\Psi_{\rm N}) \dd \vx^2 ,
\eeq
where $\Psi_{\rm N}$ is the Newtonian potential, we have
\beq
g^{\mu\nu} \pl_{\mu} t \pl_{\nu} t = \frac{-1}{1+2\Psi_{\rm N}} < 0 
\;\;\; \mbox{for} \;\;\; \Psi_{\rm N}> -1/2 .
\label{CCC-g}
\eeq
Since we focus on systems with $|\Psi_{\rm N}| \ll 1$ (e.g., $\Psi_{\rm N} \sim 10^{-6}$
in the Solar System), we have $g^{\mu\nu} \pl_{\mu} t \pl_{\nu} t <0$.
On the other hand, we obtain
\beq
G^{\mu\nu} \pl_{\mu} t \pl_{\nu} t = - \frac{\bar K' (1+2\Psi_{\rm N}) + 
\bar K'' (\pl_0\bar\varphi)^2/\cM^4}{\gamma (1+2\Psi_{\rm N})^2} ,
\eeq
whence
\beq
G^{\mu\nu} \pl_{\mu} t \pl_{\nu} t < 0 \;\; \mbox{for} \;\;
{\cal C} \equiv \bar K' + \bar K'' \frac{(\pl_0\bar\varphi)^2}{\cM^4} > 0 ,
\label{cC-def}
\eeq
where we used the approximation $1+2\Psi_{\rm N} \simeq 1$.
Around the cosmological background, where 
$\bar\chi=(\dd\bar\varphi/\dd t)^2/2\cM^4$, we obtain 
${\cal C}=\bar K'+2\bar\chi\bar K''$, whence ${\cal C}>0$.
Around a static background, we obtain ${\cal C}=\bar K'$ whence ${\cal C}>0$.

For more general backgrounds, we can see from Eq.(\ref{cC-def}) that
${\cal C}>0$ as soon as $\bar K'' \geq 0$, which for the models
(I)-(IV) of Eqs.(\ref{I-def})-(\ref{III-def}) and (\ref{cubic}) corresponds to $\bar\chi \geq 0$.
On the semiaxis $\bar\chi<0$, we have seen that ${\cal C} \simeq \bar K' > 0$
in the static limit, $| \pl\bar\varphi/\pl t | \ll | \nabla\bar\varphi |$. Therefore
the remaining case corresponds to $|\pl\bar\varphi/\pl t| \sim |\nabla\bar\varphi|$,
where time and spatial derivatives are of the same order, and with $\bar\chi<0$,
that is, $|\pl\bar\varphi/\pl t| < |\nabla\bar\varphi|$.
Then, we have
\beq
\bar\chi<0 , \;\; \bar{K''}<0 : \;\; {\cal C} \gtrsim \bar K' - | \bar K'' \bar\chi | ,
\eeq
as we assume $\bar\chi \sim - (\pl\bar\varphi/\pl t)^2/2\cM^4 
\sim - (\nabla\bar\varphi)^2 /2\cM^4$.
In the linear unscreened regime, $|\bar\chi| \ll 1$, this gives 
${\cal C} \gtrsim \bar K' \simeq 1$, whence ${\cal C} >0$. 
In the highly nonlinear screening regime, $|\bar\chi| \gg 1$ [more precisely
$\bar\chi < - \chi_*$ as in (\ref{chi*-def})], 
we have seen in Eqs.(\ref{small-K''}) and (\ref{Kss-conv}) that
$|\bar\chi \bar K''| \ll \bar K'$, because of the perihelion constraint.
Therefore, on these nonlinear scales we obtain ${\cal C} \simeq \bar K' >0$,
whether we are in the static limit, $|\pl\bar\varphi/\pl t| \ll |\nabla\bar\varphi|$,
or not, $|\pl\bar\varphi/\pl t| \sim |\nabla\bar\varphi|$, and we have already seen
that ${\cal C}>0$ when $|\pl\bar\varphi/\pl t| > |\nabla\bar\varphi|$ because
it implies $\bar\chi >0$ and $\bar K''>0$ in our models.
Note that the regime $|\pl\bar\varphi/\pl t| \sim |\nabla\bar\varphi|$ is unlikely
to occur in practice in small-scale systems, because the quasistatic approximation
applies very well, even for relatively fast matter density evolutions with 
matter flow velocities of order $v \sim c/10$ \cite{Brax:2014c}.
Therefore, we usually have $| \pl\bar\varphi/\pl t | \ll | \nabla\bar\varphi |$,
which directly gives ${\cal C} \simeq \bar K' > 0$.

Thus, we conclude that $g^{\mu\nu} \pl_{\mu} t \pl_{\nu} t <0$ and 
$G^{\mu\nu} \pl_{\mu} t \pl_{\nu} t <0$ and there are no closed causal loops
around usual astrophysical and cosmological backgrounds with $\Psi_{\rm N} > -1/2$.
This analysis fails close to neutron stars or black holes, where $\Psi_{\rm N}$ becomes
large, but this is not related to the K-mouflage model as it already appears
in the metric $g_{\mu\nu}$ in Eq.(\ref{CCC-g}).
Then, one must look for another global time coordinate, or over a large volume
around the compact object, but we leave this analysis of more extreme astrophysical 
situations to future work.

\bibliography{ref1}   

\end{document}